\documentclass[aps,pre,twocolumn,eqsecnum]{revtex4-2}
\usepackage{graphicx,color}
\usepackage{amsmath}
\usepackage{amssymb}
\usepackage{amsthm}
\usepackage{multirow}
\usepackage[normalem]{ulem}
\usepackage{hyperref}
\usepackage{enumitem}
\usepackage{bm}

\newcommand{\rme}{\mathrm{e}}

\newcommand{\Tr}{\operatorname{Tr}}

\newcommand{\diag}{\operatorname{diag}}

\newcommand{\ket}[1]{|{#1}\rangle}
\newcommand{\bra}[1]{\langle{#1}|}

\definecolor{dgreen}{rgb}{0,0.5,0}

\definecolor{dblue}{rgb}{0,0,0.6}

\definecolor{dred}{rgb}{0.784,0,0}

\definecolor{dorange}{cmyk}{0,0.72,1,0.16}
\definecolor{dmagenta}{rgb}{0.847,0.149,0.490}

\definecolor{delete}{cmyk}{0.5,0,0,0}

\begin{document}
\title{Quantum Ergotropy and Quantum Feedback Control}
\author{Kenta Koshihara}
\affiliation{Department of Physics, Waseda University, Tokyo 169-8555, Japan}
\author{Kazuya Yuasa}
\affiliation{Department of Physics, Waseda University, Tokyo 169-8555, Japan}
\date[]{June 6, 2023}
\begin{abstract}
We study the energy extraction from and charging to a finite-dimensional quantum system by general quantum operations.
We prove that the changes in energy induced by unital quantum operations are limited by the ergotropy/charging bound for unitary quantum operations. 
This implies that, in order to break the ergotropy/charging bound for unitary quantum operations, one needs to perform a quantum operation with feedback control.
We also show that the ergotropy/charging bound for unital quantum operations, applied to initial thermal equilibrium states, is tighter than the inequality representing the standard second law of thermodynamics without feedback control.
\end{abstract}
\maketitle

\section{Introduction}
\label{section:Introduction}
The energy extraction and charging are fundamental tasks in thermodynamics.
Pursuing the limitations on them leads us to the basic principles of thermodynamics.
For instance, the Gibbs states are understood to be characterized by their complete passivity (the impossibility of energy extraction from an arbitrary number of copies of them)~\cite{Pusz1978,Lenard1978,ThirringBook2002}, and the second law of thermodynamics can be phrased in terms of extractable work.
The maximum extractable work from a general nonpassive state of a quantum system via cyclic unitary operation is called ergotropy~\cite{Allahverdyan2004} and has been under intense study these years.
An interesting research direction related to the quantum ergotropy is the exploration for the enhancement of the ergotropy and the charging capacity by making use of various quantum features, such as entangling operations~\cite{Alicki2013,Hovhannisyan2013,Giorgi2015,Llobet2015PRE}, quantum correlations~\cite{Llobet2015PRX,Binder2015NJP,Salvia2022,Touli2022,Francica2022}, and quantum coherences~\cite{Francica2020,Cakmak2020}.

Another interesting feature to explore in quantum thermodynamics is quantum measurement.
The role of quantum measurement in quantum thermodynamics is not simply to acquire some information on quantum state.
Quantum measurement disturbs the state of the measured system, and thus changes the energy of the system.
In other words, quantum measurement can be used to extract/inject energy from/to a quantum working substance~\cite{Solfanelli2019}.
An extreme idea exploring this feature leads to the measurement-driven quantum heat engines~\cite{Elouard2017a,Elouard2017b,YiTalkerKim2017,Anders2017,Yi2017,Elouard2018,YiTalkerKim2018,Buffoni2019,Seah2020,Behzadi2021,Bresque2021,Anka2021,Lin2021,Manikandan2022,Lisboa2022}, where quantum measurement acts as a heat bath, fueling energy to the working substance.
Moreover, if one exploits the information acquired by a measurement, applying an additional feedback control depending on the outcome of the measurement, one would be able to extract more energy than the ergotropy and to break the bound given by the change in free energy representing the standard second law of thermodynamics without feedback control~\cite{Sagawa2008,Maruyama2009,Jacobs2009,MorikuniTasaki2011,Esposito2011,Funo2013,ParrondoHorowitzSagawa2015,Goold2016,Francica2017,Esposito2017,Manzano2018, Bernards2019,Landi2021,Koshihara2022}.

In this paper, we discuss energy extraction and charging beyond unitary operations, in particular focussing on the roles of quantum measurement and quantum feedback control.
We wish to clarify which kind of quantum operations can break the ergotropy/charging bound for unitary operations.
We give a clear answer to this question: any quantum operations without feedback control cannot break the ergotropy/charging bound for unitary operations.
This clarifies that \textit{the feedback control is necessary to break the bound for unitary operations}.
This is the main message of the paper.

There are studies on feedback strategies for the energy extraction and charging via continuous measurements in open-system scenarios, e.g.~to counteract the effects of noise~\cite{Mitchison2021} and to recover the information leaking to the environment~\cite{Morrone2023}.
Our objective is not to counteract the environmental effects, and our problem is much simpler:
we just study the change in the energy of a $d$-dimensional quantum system by a general quantum operation.
The setup of the problem is presented in Sec.~\ref{sec:Setup}\@.
The key observation is that generic quantum operations can be interpreted as quantum feedback processes.
They are indistinguishable from processes consisting of bare quantum measurements~\cite{Jacobs2009,JacobsBook2014} (or minimally disturbing quantum measurements~\cite{WisemanMilburnBook2010}) followed by quantum feedback operations.
This point is recalled in Sec.~\ref{section:QuantumOperation}\@.
We also point out that quantum operations without feedback structure are unital.
Conversely, we show that the output state $\Phi_\mathrm{unital}(\rho)$ of a unital map $\Phi_\mathrm{unital}$ for a given initial state $\rho$ can be yielded by a quantum map without feedback structure.
The unitality is thus characterized by the unnecessity of feedback control.
Then, in Sec.~\ref{section:MainResult}, we prove that the energy gain by any unital quantum operation is bounded by the ergotropy/charging bound for unitary operations.
This implies \textit{the necessity of the feedback structure in quantum operation to break the ergotropy/charging bound for unitary operations}.
As an application, in Sec.~\ref{section:Application}, we apply our bound to initial thermal equilibrium states to see an implication of our result for the second law of thermodynamics.
We show that our ergotropy/charging bound for unital operations provides a tighter bound than the standard second law of thermodynamics  without feedback control.
These results are based on the mathematical theorems proved in Appendices~\ref{appendix:UnitalIsNoFeedback}--\ref{appendix:NegativeFreeEnergiesGap}\@.

\section{setup}
\label{sec:Setup}
We are going to study how much energy one can extract from or charge to a $d$-dimensional quantum system by a general quantum operation $\Phi$.
Here, we focus on quantum operations $\Phi$ represented by completely positive and trace-preserving (CPTP) maps~\cite{Nielsen2010,Hayashi2015,ref:DynamicalMap-Alicki}; we do not consider trace-decreasing maps, such as quantum operations with postselections.
Suppose that before a quantum operation $\Phi$ the system is in a state $\rho$ with a Hamiltonian $H$.
By the quantum operation $\Phi$, the state of the system is changed from $\rho$ to $\rho'$,
\begin{equation}
\rho
\to
\rho'=\Phi(\rho),
\end{equation}
during which the Hamiltonian may be steered from $H$ to $H'$.
The gain of the energy by the quantum operation $\Phi$ is given by 
\begin{equation}
\Delta E(\rho)
=
\Tr(H'\rho')
-\Tr(H\rho).
\label{eqn:EnergyGainDef}
\end{equation}
We wish to clarify the bound on $\Delta E(\rho)$, i.e., the maximum extractable/chargeable energy, by general quantum operations $\Phi$, in particular focusing on the roles of quantum measurement and quantum feedback control.

\section{Structure of Quantum Operation}
\label{section:QuantumOperation}
\subsection{General Quantum Operations and Feedback Control}
\label{subsection:PolarDecomposition}
Before starting to investigate the energy extraction and charging, let us recall the structure of general quantum operation.
The quantum operations of the most basic kind are the unitary operations,
\begin{equation}
\Phi_U(\rho)=U\rho U^\dag,
\label{eq:Unitary}
\end{equation}
represented by unitary operators $U$, satisfying $U^\dag U=\openone$.
In quantum mechanics, however, one can think of more general quantum operations.
General quantum operation $\Phi$ (without postselection) is described by a CPTP map, and is given by the Kraus representation~\cite{ref:DynamicalMap-Alicki,Nielsen2010,Hayashi2015}
\begin{equation}
\Phi(\rho)=\sum_i K_i\rho K_i^\dag,
\label{eq:KrausRepresentation}
\end{equation}
with $\{K_i\}$ a set of Kraus operators satisfying $\sum_i K_i^\dag K_i=\openone$. 
Interesting and relevant examples of nonunitary operation include quantum measurements.
They provoke nonunitary changes in quantum state and are regarded as quantum operations~\cite{Nielsen2010,Hayashi2015}.
If one accepts any outcomes of a quantum measurement without postselection (nonselective measurement), the net effect of the quantum measurement on quantum state is described by a CPTP map~(\ref{eq:KrausRepresentation}).
Feedback control, where an additional unitary operation is applied depending on the outcome of the measurement, also results in a nonunitary transformation on quantum state.

\begin{figure}[t]
\includegraphics[width=0.9\linewidth]{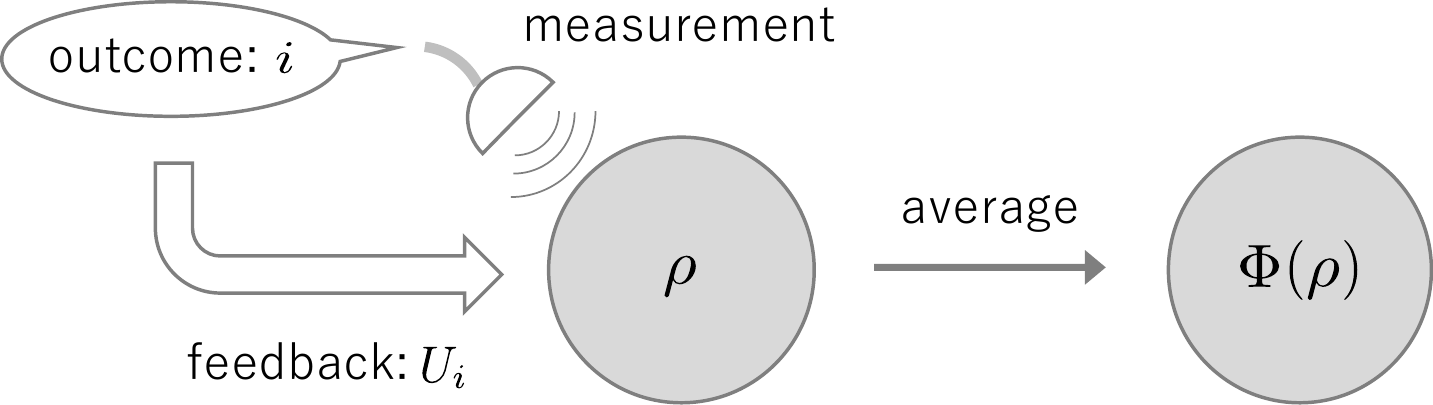}
\caption{Quantum operation $\Phi$ as a quantum feedback control.}
\label{fig:FeedbackControl}
\end{figure}

Generic quantum operations represented by the Kraus representation~(\ref{eq:KrausRepresentation}) are actually regarded as (or indistinguishable from) quantum feedback controls.
To see this, consider the polar decomposition of each Kraus operator $K_i$ of the general quantum operation $\Phi$ in~(\ref{eq:KrausRepresentation}),
\begin{equation}
K_i=U_i M_i,
\quad
M_i\ge0,
\label{eq:PolarDecomposition}
\end{equation}
where $M_i$ is a positive-semidefinite operator and $U_i$ is a unitary~\cite{Nielsen2010,WisemanMilburnBook2010,HornJohnson2012,JacobsBook2014,Hayashi2015}. 
Then, the map~(\ref{eq:KrausRepresentation}) is written as
\begin{equation}
\Phi(\rho)
=
\sum_i U_iM_i\rho M_iU_i^\dag,
\label{eq:FeedbackControl}
\end{equation}
and can be interpreted as a quantum feedback process, where  a measurement represented by the set of measurement operators $\{M_i\}$ is performed and then a unitary control $U_i$ is applied depending on the outcome $i$ of the measurement~\cite{Jacobs2009,JacobsBook2014,Koshihara2022}.
See Fig.~\ref{fig:FeedbackControl}.
Note that the normalization condition on the Kraus operators $\{K_i\}$ is translated to $\sum_i M_i^2=\openone$, and the operators $\{M_i\}$ are valid measurement operators.
We will see that the presence of such a feedback structure in a quantum operation $\Phi$, whether or not it is intentionally implemented by an experimenter, makes a difference in the maximum extractable or chargeable energy on a quantum system: \emph{the feedback structure is necessary to go beyond the ergotropy/charging bound for unitary operations}.

\subsection{Quantum Operations without Feedback Control}
\label{subsection:WithoutFeedbackControl}
The quantum measurement process
\begin{equation}
\Phi_\mathrm{ms}(\rho)
=
\sum_i M_i\rho M_i
\label{eq:BareMeasurement}
\end{equation}
represented by the positive-semidefinite measurement operators $\{M_i\}$, with unitaries removed from the Kraus operators, is called ``bare'' quantum measurement \cite{Jacobs2009,JacobsBook2014} or ``minimally disturbing'' quantum measurement \cite{WisemanMilburnBook2010}.
It is considered as a ``pure'' quantum measurement, without any feedback operations involved.
Projective measurements, with the measurement operators $\{M_i\}$ given by orthogonal projections $\{P_i\}$ satisfying $P_i=P_i^\dag$, $P_iP_j=P_i\delta_{ij}$, and $\sum_iP_i=\openone$, are the canonical examples of the bare quantum measurement.

Even if an additional unitary operation $U$ is applied after such a bare quantum measurement as
\begin{equation}
(\Phi_U\circ\Phi_\mathrm{ms})(\rho)
=
\sum_i UM_i\rho M_i U^\dag,
\label{eq:Composite}
\end{equation}
it is not considered as a feedback control, since the unitary $U$ is just applied irrespective of the outcome $i$ of the measurement.
We point out that the absence of the feedback structure in quantum operation is closely related to the unitality of operation.
A quantum operation $\Phi_\mathrm{unital}$ is called unital if it preserves the identity,
\begin{equation}
\Phi_\mathrm{unital}(\openone)
=
\openone.
\label{eq:Unitality}
\end{equation}
Any quantum operations of the type~(\ref{eq:Composite}), without feedback structure, are unital, satisfying the unitality condition~(\ref{eq:Unitality}).
Its contrapositive says that, if a quantum operation $\Phi$ is not unital, $\Phi(\openone)\neq\openone$, then it is for sure endowed with a feedback structure as~(\ref{eq:FeedbackControl}), involving different unitaries $\{U_i\}$ in the Kraus operators.
The converse is not necessarily true: there exist unital maps $\Phi_\mathrm{unital}$ that are endowed with feedback structures. 
However, the following variant of the converse is true: given a unital map $\Phi_\mathrm{unital}$ and an input state $\rho$, there exist a set of orthogonal rank-1 projections $\{P_i\}$ and a unitary $U$ such that~\cite{footnote:UnitalFeedback}
\begin{equation}
\Phi_\mathrm{unital}(\rho)
=
\sum_{i=1}^d U P_i\rho P_i U^\dag.
\label{eq:UnitalIsNoFeedback}
\end{equation}
We prove it in Appendix~\ref{appendix:UnitalIsNoFeedback}\@.
This means that, for each input state $\rho$, the same effect as the one induced by a unital operation $\Phi_\mathrm{unital}$ can be achieved without feedback control.
In this way, the unitality of an operation $\Phi_\mathrm{unital}$ is characterized by the unnecessity of feedback control.

\section{Necessity of Feedback Control to Break the Ergotropy Bound}
\label{section:MainResult}
Let us now investigate the energy extraction and charging by general quantum operations.

\subsection{Generalized Ergotropy/Charging Bound}
\label{subsection:GeneralizedErgotropyBound}
We first review the ergotropy bound and provide a generalization of it.
In its original formulation~\cite{Allahverdyan2004}, ergotropy refers to the maximum extractable energy via unitary operation $\Phi_U$ realized by driving the Hamiltonian of the working substance in a cyclic manner.
Let $\Delta E_U(\rho)$ be the energy gain~(\ref{eqn:EnergyGainDef}) obtained by performing a unitary operation $\Phi_U$ on the working substance in the state $\rho$.
After the operation, the driven Hamiltonian is brought back to the initial one $H$.
Then, one can prove that the following lower bound on $\Delta E_U(\rho)$ holds,
\begin{equation}
\Delta E_{U}(\rho) \ge -\mathcal{E}_{\rho}^-
\quad
\mathrm{if}
\quad
H\to H,
\label{eq:ErgotropyBound}
\end{equation}
with
\begin{equation}
\mathcal{E}_\rho^-=\Tr(H\rho)-\bm{\varepsilon}^\uparrow\cdot\bm{r}^\downarrow,
\label{eq:ErgotropyMinus}
\end{equation}
where $\bm{\varepsilon}^\uparrow$ is the $d$-dimensional vector consisting of the eigenvalues of the Hamiltonian $H$ arranged in the increasing order $\varepsilon_{1}^{\uparrow}\le\varepsilon_{2}^{\uparrow}\le\cdots\le\varepsilon_{d}^{\uparrow}$, and $\bm{r}^{\downarrow}$ is the $d$-dimensional vector consisting of the eigenvalues of the initial density operator $\rho$ arranged in the decreasing order $r_{1}^{\downarrow}\ge r_{2}^{\downarrow}\ge\cdots\ge r_{d}^{\downarrow}$.
In the following, $\bm{v}^{{\uparrow}(\downarrow)}$ of a vector $\bm{v}$ means the vector obtained by rearranging the elements of $\bm{v}$ in the increasing (decreasing) order.
The quantity $\mathcal{E}_{\rho}^-\,(\ge0)$ in~(\ref{eq:ErgotropyMinus}) is always nonnegative (see Appendix~\ref{appendix:GeneralizedErgotropyBound}), giving the maximum extractable energy from the system, and it is called ergotropy.
It depends on the initial state $\rho$ and the initial/final Hamiltonian $H$~\cite{footnote:Unknown}.
This bound is achievable by a suitable unitary operation $\Phi_U$.

As noted in Ref.~\cite{Binder2015NJP}, one can also derive the upper bound on $\Delta E_U(\rho)$, i.e., the maximum chargeable energy on the system.
Moreover, one can generalize the bound to accommodate the situations where the driven Hamiltonian is not brought back to the initial one $H$ but is left $H'$ at the end of the unitary operation $\Phi_U$.
The bound valid for such a generalized situation reads
\begin{equation}
\Delta\bm{\varepsilon}^\uparrow\cdot\bm{r}^\downarrow
-
\mathcal{E}^-_\rho
\le
\Delta E_U(\rho)
\le
\Delta\bm{\varepsilon}^\uparrow\cdot\bm{r}^\uparrow
+\mathcal{E}^+_\rho,
\label{eq:GeneralizedErgotropyBound}
\end{equation}
with 
\begin{equation}
\mathcal{E}_\rho^+=\bm{\varepsilon}^\uparrow\cdot\bm{r}^\uparrow-\Tr(H\rho),
\label{eq:ErgotropyPlus}
\end{equation}
where $\Delta\bm{\varepsilon}^\uparrow=\bm{\varepsilon}^{\prime\uparrow}-\bm{\varepsilon}^\uparrow$, with $\bm{\varepsilon}^{\prime\uparrow}$ consisting of the eigenvalues of the final Hamiltonian $H'$ in the increasing order.
The quantity $\mathcal{E}_{\rho}^+\,(\ge0)$ is also always nonnegative (see Appendix~\ref{appendix:GeneralizedErgotropyBound}), and gives the upper bound on $\Delta E_U(\rho)$, i.e., the maximum chargeable energy on the system, when the Hamiltonian is brought back to the initial one $H'=H$.
We provide the proof of the generalized bound~(\ref{eq:GeneralizedErgotropyBound}) in Appendix~\ref{appendix:GeneralizedErgotropyBound}\@.
Both lower and upper bounds of~(\ref{eq:GeneralizedErgotropyBound}) are tight, and can be reached by suitable unitary operations $\Phi_U$.

\subsection{Ergotropy/Charging Bound for Unital Quantum Operations}
\label{subsection:UnitalErgotropyBound}
We now present the key result of this paper.
Let $\Delta E_\mathrm{unital}(\rho)$ be the energy gain~(\ref{eqn:EnergyGainDef}) obtained by performing a unital quantum operation $\Phi_\mathrm{unital}$ on the working substance in the state $\rho$.
The initial Hamiltonian of the working substance is $H$, which may be changed to $H'$ at the end of the unital operation $\Phi_\mathrm{unital}$.
Then, the following bound on $\Delta E_\mathrm{unital}(\rho)$ holds,
\begin{equation}
\Delta\bm{\varepsilon}^\uparrow\cdot\bm{r}^\downarrow
-
\mathcal{E}^-_\rho
\le
\Delta E_\mathrm{unital}(\rho)
\le
\Delta\bm{\varepsilon}^\uparrow\cdot\bm{r}^\uparrow
+\mathcal{E}^+_\rho
.
\label{eq:UnitalErgotropyBound}
\end{equation}
This is exactly the same as the ergotropy/charging bound presented in~(\ref{eq:GeneralizedErgotropyBound}) for the energy gain $\Delta E_U(\rho)$ obtained by a unitary operation $\Phi_U$.
Note that both lower and upper bounds of~(\ref{eq:UnitalErgotropyBound}) are tight.
In fact, unitary operations $\Phi_U$ are special instances of unital operation $\Phi_\mathrm{unital}$, and the lower and upper bounds of~(\ref{eq:UnitalErgotropyBound}) are achieved by the unitary operations $\Phi_U$ saturating the lower and upper bounds of~(\ref{eq:GeneralizedErgotropyBound}).

The bound~(\ref{eq:UnitalErgotropyBound}) shows that any unital operations $\Phi_\mathrm{unital}$ cannot extract or charge energy beyond the ergotropy/charging bound~(\ref{eq:GeneralizedErgotropyBound}) for unitary operations $\Phi_U$.
This implies, for instance, that the energy change induced by the backaction of a bare quantum measurement $\Phi_\mathrm{ms}$ is limited by the ergotropy/charging bound~(\ref{eq:GeneralizedErgotropyBound}) (cf., the energy extraction by projective measurement is studied in Ref.~\cite{Solfanelli2019}).
More interestingly, since the nonunitality requires some feedback structure in the operation, as clarified in Sec.~\ref{subsection:WithoutFeedbackControl}, the bound~(\ref{eq:UnitalErgotropyBound}) reveals that, \emph{in order to extract or charge energy beyond the ergotropy/charging bound~(\ref{eq:GeneralizedErgotropyBound}), a feedback control is necessary}.
This is the main message of this paper.
It is indeed possible to break the ergotropy/charging bound~(\ref{eq:GeneralizedErgotropyBound}) via a feedback control, as we will see later with some simple examples.

Let us prove the bound~(\ref{eq:UnitalErgotropyBound}).
We just have to recall Uhlmann's representation theorem for unital maps $\Phi_\mathrm{unital}$: given a unital map $\Phi_\mathrm{unital}$ and an input state $\rho$, there exist a probability distribution $\{p_i\}$ and a set of unitaries $\{U_i\}$ such that
\begin{equation}
\Phi_\mathrm{unital}(\rho)
=
\sum_{i=1}^{d!}p_i U_i\rho U_i^\dag
.
\label{eq:UnitalUhlmann}
\end{equation}
See, e.g., Refs.~\cite{Nielsen2001,Nielsen2010,Sagawa2022}.
This means that, for each input state $\rho$, the same output state $\Phi_\mathrm{unital}(\rho)$ as the one yielded by the unital map $\Phi_\mathrm{unital}$ can be obtained by a random unitary operation as~(\ref{eq:UnitalUhlmann})~\cite{footnote:RandomUnitary,footnote:MarcusReeTheorem}.
This provides another characterization of the unitality of an operation $\Phi_\mathrm{unital}$, alternative to the characterization presented in~(\ref{eq:UnitalIsNoFeedback}). 
Using this representation~(\ref{eq:UnitalUhlmann}), one can bound the energy gain $\Delta E_\mathrm{unital}(\rho)$ obtained by a unital quantum operation $\Phi_\mathrm{unital}$ as
\begin{align}
\Delta E_\mathrm{unital}(\rho)
={}&
\Tr[H'\Phi_\mathrm{unital}(\rho)]
-\Tr(H\rho)
\vphantom{\sum_i}
\nonumber\\
={}&
\sum_ip_i\Tr(H' U_i\rho U_i^\dag)
-\Tr(H\rho)
\nonumber\\
\le{}&
\sum_ip_i\max_U\Tr(H' U\rho U^\dag)
-\Tr(H\rho)
\nonumber\\
={}&
\max_U\Tr(H' U\rho U^\dag)
-\Tr(H\rho)
\vphantom{\sum_i}
\nonumber\\
={}&
\max_U\Delta E_U(\rho)
\vphantom{\sum_i}
\nonumber\\
={}&
\Delta\bm{\varepsilon}^\uparrow\cdot\bm{r}^\uparrow
+\mathcal{E}^+_\rho.
\label{eq:UnitalLowerBound}
\end{align}
This proves the upper bound of~(\ref{eq:UnitalErgotropyBound}).
The last equality is due to the upper bound of~(\ref{eq:GeneralizedErgotropyBound}) for unitary operations $\Phi_U$.
The lower bound of~(\ref{eq:UnitalErgotropyBound}) is proven in the same way, but $\max_U$ in~(\ref{eq:UnitalLowerBound}) is replaced by $\min_U$ and the lower bound of~(\ref{eq:GeneralizedErgotropyBound}) is used instead of the upper bound.

\subsection{Energetic Advantage of Feedback Control}
\label{subsection:NumericalExample1}
Let us look at a few simple examples.
We first consider a three-level system, which is given in an initial state $\rho$ with an initial Hamiltonian $H$.
We perform a quantum operation $\Phi$ on the system, and see how the energy of the system is changed from $E(\rho)=\Tr(H\rho)$ to $E\bm{(}\Phi(\rho)\bm{)}=\Tr[H\Phi(\rho)]$.
Here, we assume that the Hamiltonian of the system is brought back to the initial one $H$ after the operation $\Phi$.
If the quantum operation $\Phi$ is not unitary, it also changes the entropy of the system in general.
We therefore look at the change of the von Neumann entropy from $S(\rho)=-\Tr(\rho\ln\rho)$ to $S\bm{(}\Phi(\rho)\bm{)}$ as well.
We display those quantities on the energy-entropy diagram~\cite{Sparaciari2017,Bera2019}.

\begin{figure}[b]
\begin{tabular}{l@{\ }l}
\footnotesize(a)
&
\footnotesize(b)
\\
\includegraphics[height=0.318\linewidth]{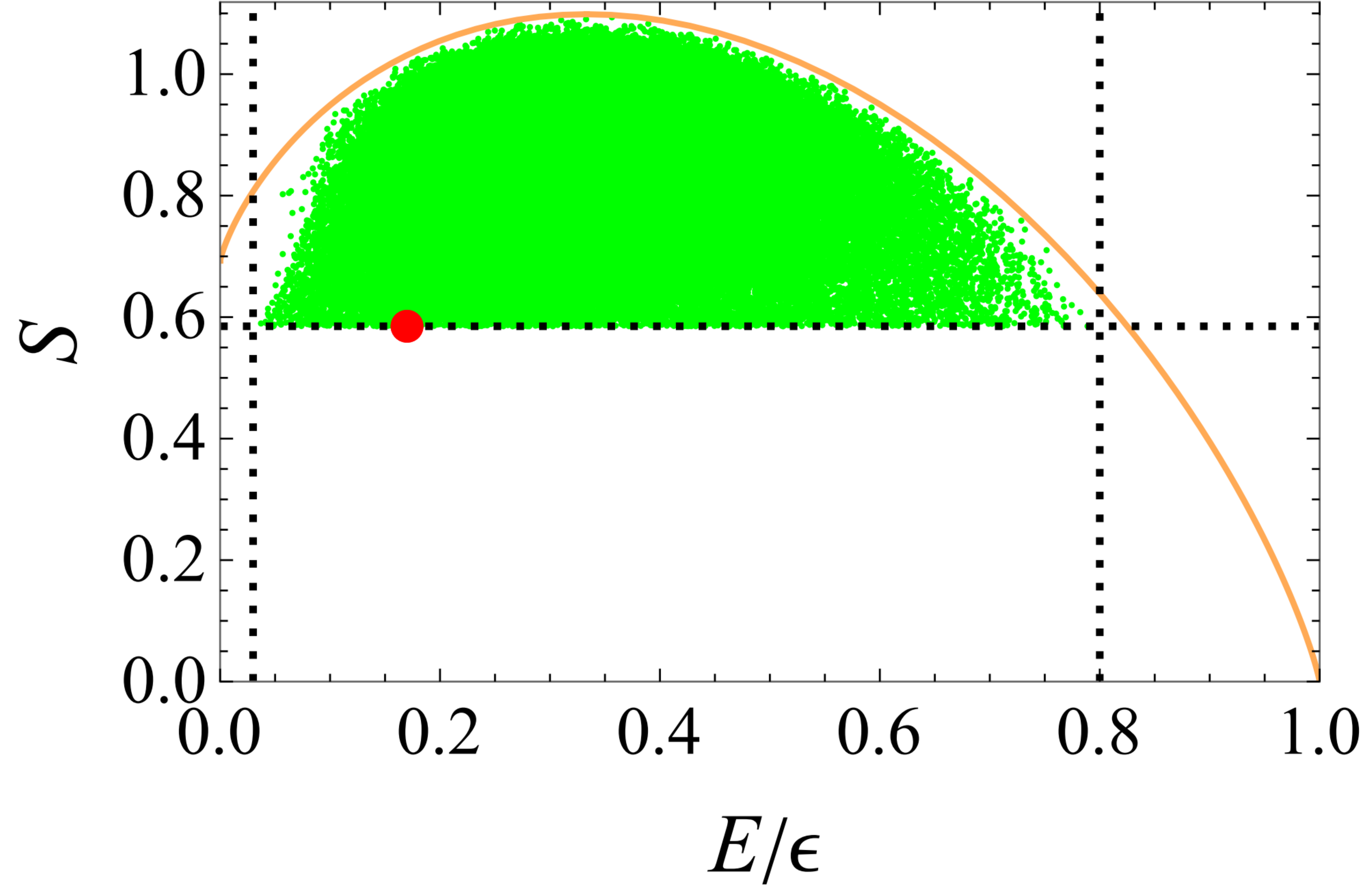}
&
\includegraphics[height=0.318\linewidth]{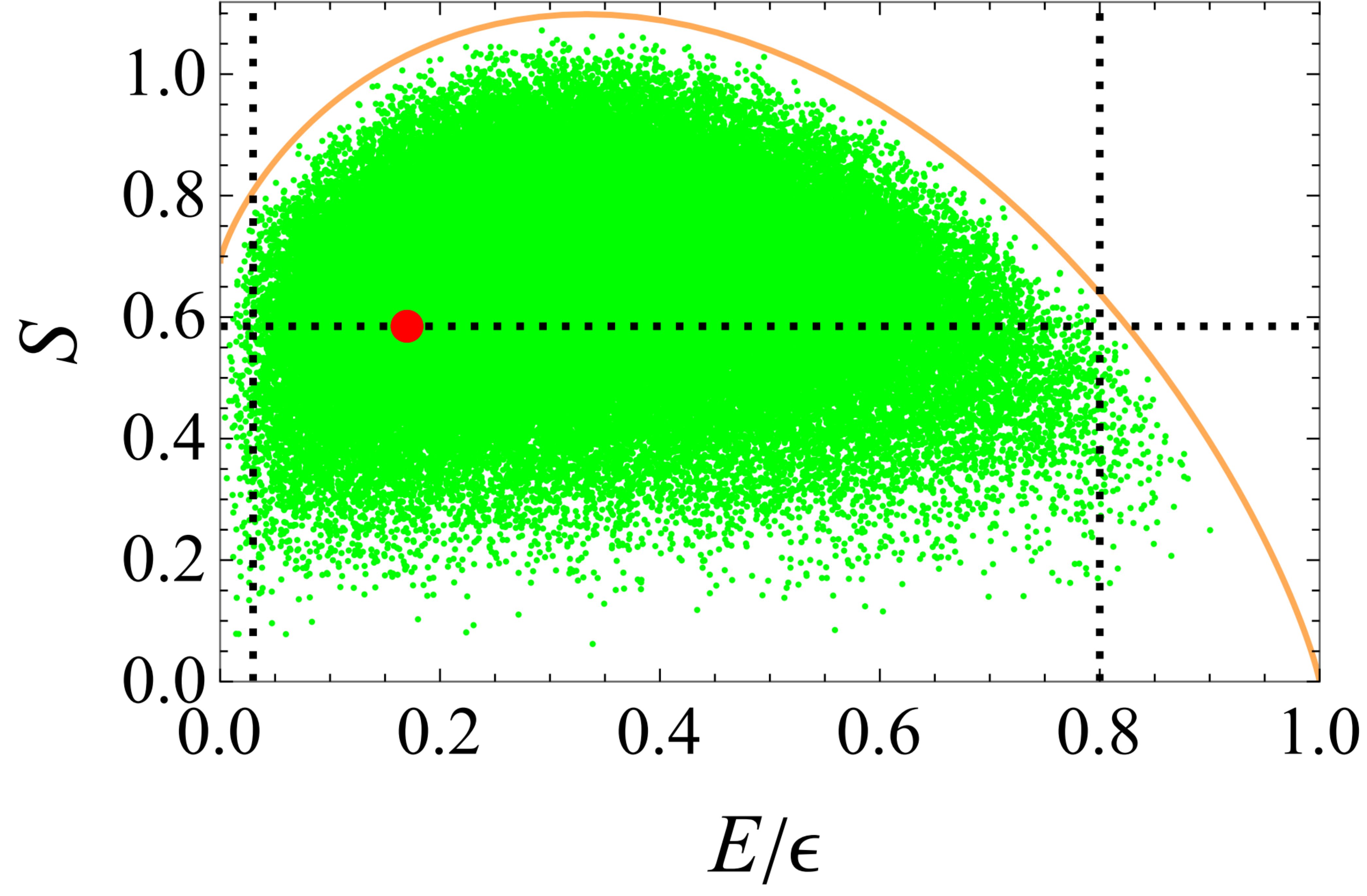}
\end{tabular}
\caption{The energy-entropy diagram for the initial state $\rho=\diag(0.8,0.03,0.17)$ and the initial/final Hamiltonian $H=\diag(0,0,\epsilon)$ of a three-level system.
The initial energy-entropy point is indicated by the red filled circle.
The energy-entropy points of the output states of unital operations $\Phi_\mathrm{unital}$ are shown by the green dots in (a), while those of nonunital operations $\Phi_\mathrm{fb}$, endowed with feedback structures, are shown in (b).
See the main text concerning how the unital and nonunital operations are sampled.
The left and right vertical dotted lines are given by $\bm{\varepsilon}^\uparrow\cdot\bm{r}^\downarrow$ and $\bm{\varepsilon}^\uparrow\cdot\bm{r}^\uparrow$, indicating the ergotropy and charging bounds of~(\ref{eq:UnitalErgotropyBound}), respectively.
The solid orange concave curve is given by the Gibbs states $\rho_\beta=\rme^{-\beta H}/Z_\beta$ with $\beta\in\mathbb{R}$, and shows the upper bound on the entropy.
In particular, the tip of the concave curve represents the maximally mixed state $\rho_0=\openone/3$ with $\beta=0$.
Because of the degeneracy in the lower energy eigenvalue of $H$, there exist mixed states along the low-energy vertical border.
The horizontal dotted line is equientropic to the initial state.}
\label{fig:ESdiagram1}
\end{figure}

In Fig.~\ref{fig:ESdiagram1}, we consider the initial state $\rho=\diag(0.8,0.03,0.17)$ and the initial/final Hamiltonian $H=\diag(0,0,\varepsilon)$.
The initial pair of the energy and the entropy is displayed by the filled circle on the energy-entropy plane.
In Fig.~\ref{fig:ESdiagram1}(a), the energy-entropy pairs after unital operations $\Phi_\mathrm{unital}$ for the given initial state $\rho$ and the given initial/final Hamiltonian $H$ are shown by the dots, where the unital operations $\Phi_\mathrm{unital}$ are sampled on the basis of Uhlmann's representation~(\ref{eq:UnitalUhlmann}), with the probability distribution $\{p_i\}_{i=1,\ldots,3!}$ and the unitaries $\{U_i\}_{i=1,\ldots,3!}$ chosen randomly and uniformly.
By unital operations $\Phi_\mathrm{unital}$, one cannot extract or charge energy beyond the ergotropy/charging bound~(\ref{eq:UnitalErgotropyBound}).
Indeed, the sampled dots are confined between the two vertical lines showing the ergotropy and charging bounds.
While unitary operations $\Phi_U$ preserve the entropy, unital operations $\Phi_\mathrm{unital}$ generally increase it, $\Delta S_\mathrm{unital}=S\bm{(}\Phi_\mathrm{unital}(\rho)\bm{)}-S(\rho)\ge0$ [see Eq.~(5.18) of Ref.~\cite{Sagawa2022} and Corollary~7.10 of Ref.~\cite{HolevoBook2019}].
On the other hand, for each fixed energy $E$, the maximum entropy is reached by the Gibbs state $\rho_\beta=\rme^{-\beta H}/Z_\beta$, with $\beta\in\mathbb{R}$ determined by $E$.
This is nothing but the principle of maximum entropy~\cite{Sparaciari2017}.
This upper bound on the entropy is shown by the concave curve on the energy-entropy plane.
The energy-entropy pairs of the output states of unital operations $\Phi_\mathrm{unital}$ are thus confined within the region whose boundary is made by the ergotropy and charging bounds, the equientropic line, and the Gibbs states~\cite{footnote:UnitalSpaces}.

\begin{figure}[t]
\begin{tabular}{l@{\ }l}
\footnotesize(a)
&
\footnotesize(b)
\\
\includegraphics[height=0.318\linewidth]{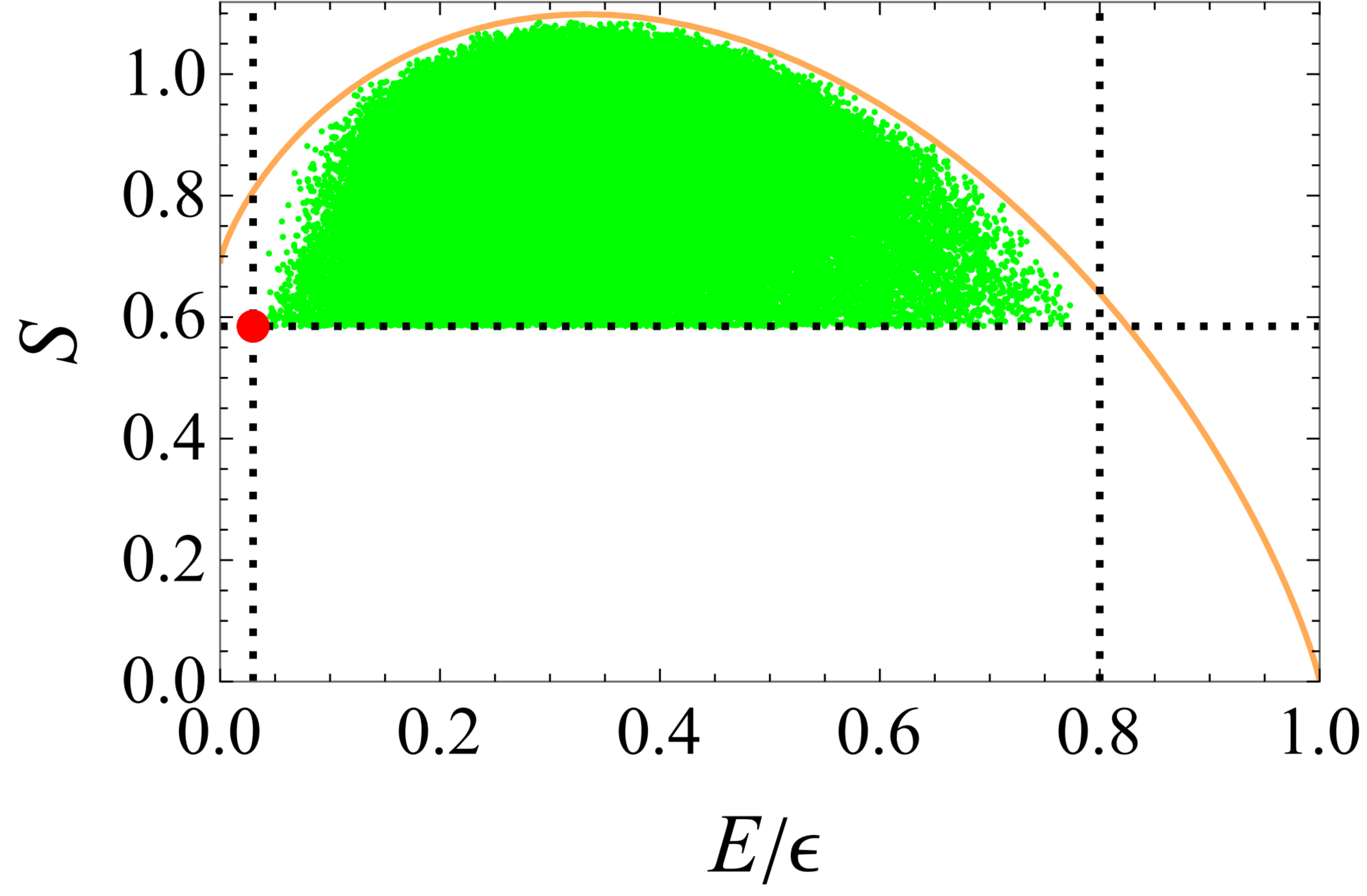}
&
\includegraphics[height=0.318\linewidth]{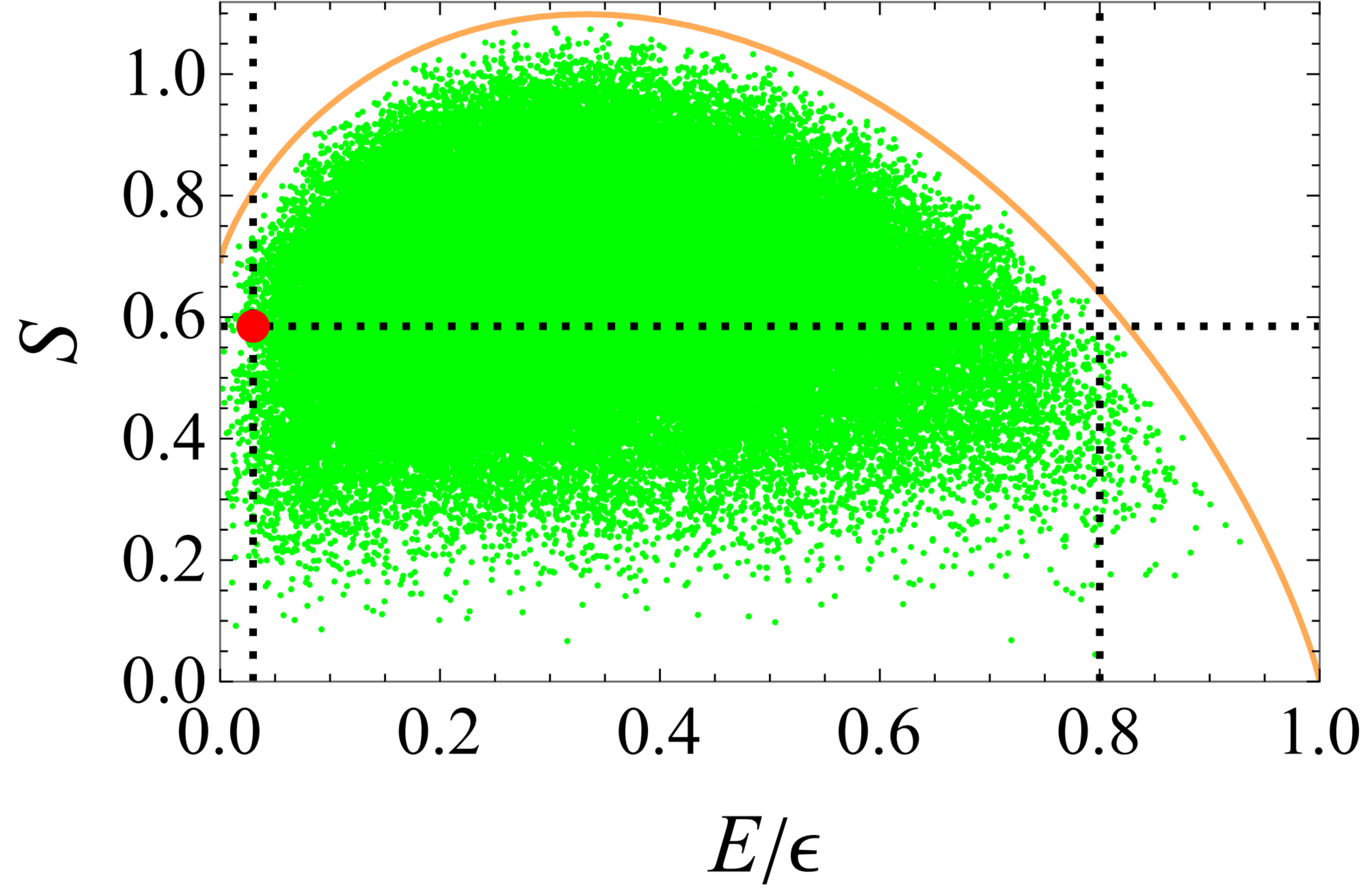}
\end{tabular}
\caption{The same as Fig.~\ref{fig:ESdiagram1}, but the initial state is replaced by $\rho=\diag(0.8,0.17,0.03)$.
(a) for unital operations $\Phi_\mathrm{unital}$ and (b) for nonunital operations $\Phi_\mathrm{fb}$.
Since this initial state $\rho$ is a passive state, the vertical dotted line indicating the ergotropy bound $\bm{\varepsilon}^\uparrow\cdot\bm{r}^\downarrow$ runs through the initial point shown by the red filled circle.}
\label{fig:ESdiagram2}
\end{figure}

In Fig.~\ref{fig:ESdiagram1}(b), on the other hand, the energy-entropy pairs of the output states of nonunital operations $\Phi_\mathrm{fb}$, which are generically endowed with feedback structures, are shown by the dots, for the same initial state $\rho$ and the same initial/final Hamiltonian $H$ as those taken in Fig.~\ref{fig:ESdiagram1}(a).
Here, the nonunital operations $\Phi_\mathrm{fb}$ are sampled on the basis of the formula~(\ref{eq:FeedbackControl}): the measurement operators $\{M_i\}_{i=1,2,3}$ are constrained to rank-1 orthogonal projections $M_i=\ket{\psi_i}\bra{\psi_i}$ ($i=1,2,3$), and their basis vectors $\{\ket{\psi_i}\}_{i=1,2,3}$ and the feedback unitaries $\{U_i\}_{i=1,2,3}$ are chosen randomly and uniformly.
It is clear from the plot that one can extract/charge energy beyond the ergotropy/charging bound, in the presence of a feedback structure in the operation $\Phi_\mathrm{fb}$.
In addition, we see that nonunital operations $\Phi_\mathrm{fb}$, endowed with feedback structures, can decrease the entropy, in contrast to unital operations $\Phi_\mathrm{unital}$, which can only increase the entropy~\cite{HolevoBook2019,Sagawa2022}.
Actually, any points in the region between the concave curve, representing the upper bound on the entropy, and the zero-entropy line, which is the lower bound, are reachable from any state by general quantum operations $\Phi_\mathrm{fb}$~\cite{footnote:WholeRegion}.
Because of the concavity of the maximum-entropy curve, in order to extract or charge a large amount of energy, it is helpful to decrease the entropy by a feedback control.
In particular, the lowest (highest) energy state can be reached from any initial state, achieving the maximum energy extraction (charging), by first performing a projective measurement to collapse the state into a pure state $\ket{\psi_i}$, and then transforming it into the lowest (highest) energy state by an additional unitary feedback control. 
The energy gain $\Delta E(\rho)$ by a general nonunital operation $\Phi_\mathrm{fb}$, where the Hamiltonian of the system is steered from the initial one $H$ to the final one $H'$ in general, is thus bounded by
\begin{equation}
\varepsilon_1'^\uparrow-E(\rho)\le \Delta E(\rho)\le\varepsilon_d'^\uparrow-E(\rho),
\label{eq:NonunitalBound}
\end{equation}
where $\varepsilon_1'^\uparrow$ and $\varepsilon_d'^\uparrow$ are the lowest and highest energy eigenvalues of the final Hamiltonian $H'$, respectively.
Both lower and upper bounds are reachable.

\begin{figure}[t]
\begin{tabular}{l@{\ }l}
\footnotesize(a)
&
\footnotesize(b)
\\
\includegraphics[height=0.318\linewidth]{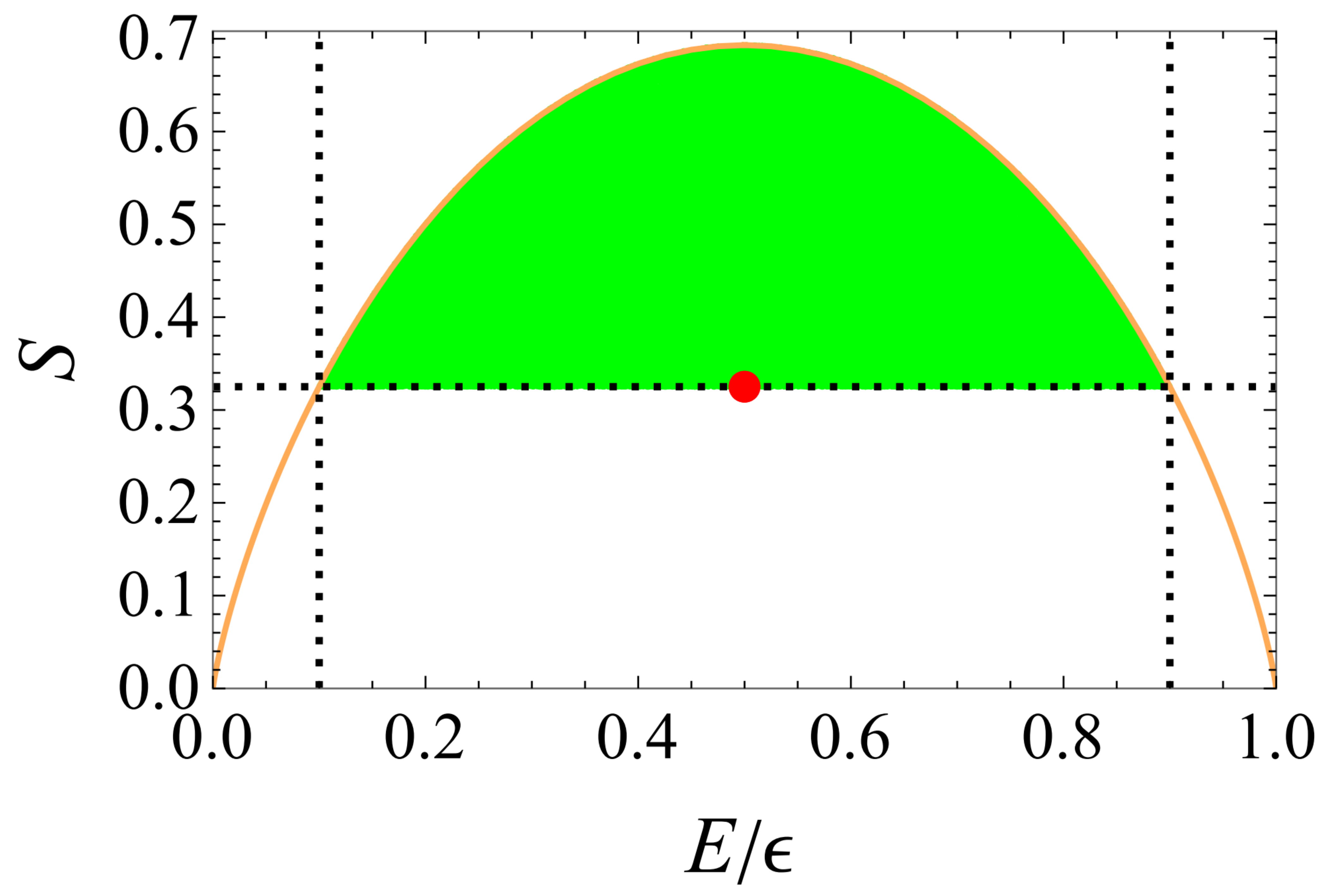}
&
\includegraphics[height=0.318\linewidth]{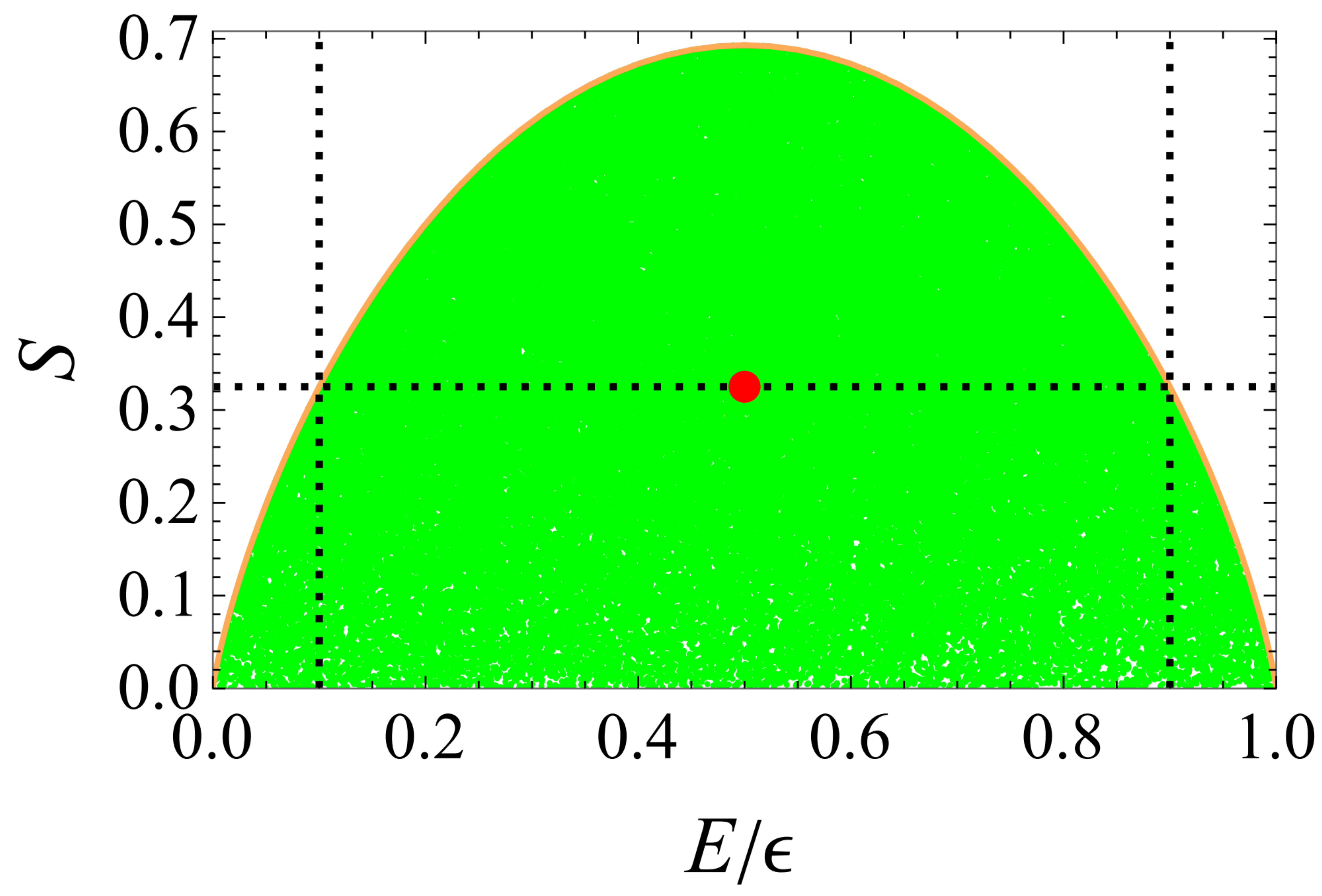}
\end{tabular}
\caption{The energy-entropy diagram for the initial state $\rho=\left(\begin{smallmatrix}0.5&-0.4\\-0.4&0.5\end{smallmatrix}\right)$ and the initial/final Hamiltonian $H=\diag(0,\epsilon)$ of a two-level system. (a) for unital operations $\Phi_\mathrm{unital}$ and (b) for nonunital operations $\Phi_\mathrm{fb}$.
The initial energy-entropy point is indicated by the red filled circle.
The vertical dotted lines indicating the ergotropy and charging bounds always intersect with the horizontal dotted equientropic line on the solid orange concave maximum-entropy curve for any two-level system, since a Gibbs state $\rho_\beta$ is reachable from any initial state $\rho$ by a unitary operation $\Phi_U$.}
\label{fig:ESdiagram4}
\end{figure}

In Fig.~\ref{fig:ESdiagram2}, the initial state is replaced by $\rho=\diag(0.8,0.17,0.03)$, while the initial/final Hamiltonian is the same as the one in Fig.~\ref{fig:ESdiagram1}, i.e., $H=\diag(0,0,\varepsilon)$.
In this case, the ergotropy is vanishing and no energy is extractable from the given initial state $\rho$ without feedback control.
This is due to the passivity of the initial state $\rho$~\cite{Pusz1978,Lenard1978,ThirringBook2002}.
A state is called passive, if lower energy states are more occupied, and the initial state $\rho$ chosen here satisfies this condition.
It is known that no energy is extractable from a passive state by a cyclic unitary operation $\Phi_U$.
It is also the case for unital operations $\Phi_\mathrm{unital}$~\cite{Anka2021}.

The energy-entropy diagram looks much simpler for two-level systems.
See Fig.~\ref{fig:ESdiagram4}.
The concave maximum-entropy curve is always symmetric for two-level systems.
In addition, the vertical ergotropy/charging lines and the horizontal equientropic line always intersect on the concave maximum-entropy curve.
This is because a Gibbs state $\rho_\beta$ is reachable from any initial state $\rho$ by a unitary operation $\Phi_U$ for a two-level system.

\section{Application to Equilibrium Initial State}
\label{section:Application}
Let us focus on systems initially in thermal equilibrium, to see how our main result~(\ref{eq:UnitalErgotropyBound}) is relevant to the second law of thermodynamics.

\subsection{Second Law of Thermodynamics}
\label{subsection:SecondLaw}
Let us first recall a generalized version of the second law of thermodynamics~\cite{Sagawa2008,Esposito2011}.
Consider the thermal equilibrium state $\rho_\beta=\rme^{-\beta H}/Z_\beta$ at an inverse temperature $\beta\,(>0)$. 
After some operation, the state is changed from $\rho_\beta$ to $\rho'$, and the Hamiltonian is changed from $H$ to $H'$.
A generalized version of the second law of thermodynamics~\cite{Sagawa2008,Esposito2011} says that the energy gain $\Delta E(\rho_\beta)=\Tr(H'\rho')-\Tr(H\rho_\beta)$ from the thermal equilibrium state $\rho_\beta$ is bounded by
\begin{equation}
\Delta E(\rho_\beta)
\ge
\Delta F_\beta
+\frac{1}{\beta}\Delta S(\rho_\beta),
\label{eq:Generalized2ndLaw}
\end{equation}
where $\Delta S(\rho_\beta)=S(\rho')-S(\rho_\beta)$ is the change in the von Neumann entropy from the thermal equilibrium state $\rho_\beta$, and $\Delta F_\beta=F_\beta'-F_\beta$ is the difference  between the free energies $F_\beta=-\beta^{-1}\log Z_\beta$ and $F'_\beta=-\beta^{-1}\log Z'_\beta$ defined through the partition functions $Z_\beta=\Tr\rme^{-\beta H}$ and $Z_\beta'=\Tr\rme^{-\beta H'}$ based on the initial and final Hamiltonians $H$ and $H'$, respectively. 
The inequality~(\ref{eq:Generalized2ndLaw}) is a direct consequence of the inequality between the equilibrium and nonequilibrium free energies [Eq.~(11) of Ref.~\cite{Esposito2011}, Eq.~(42) of Ref.~\cite{Landi2021}, and Eq.~(8.5) of Ref.~\cite{Sagawa2022}],
\begin{equation}
f_\beta(\rho,H)\ge F_\beta,
\label{eq:FreeEnergiesGap}
\end{equation}
valid for any density operator $\rho$, any Hamiltonian $H$, and any positive inverse temperature $\beta\,(>0)$.
While the equilibrium free energy $F_\beta$ is defined through the partition function $Z_\beta$ as above, the nonequilibrium free energy $f_\beta(\rho,H)$ is defined for an arbitrary (generally nonequilibrium) state $\rho$ by $f_\beta(\rho,H)=\Tr(H\rho)-S(\rho)/\beta$. 
The equality $f_\beta(\rho,H)=F_\beta$ holds if and only if $\rho=\rho_\beta=\rme^{-\beta H}/Z_\beta$.
The inequality~(\ref{eq:Generalized2ndLaw}) is obtained immediately from the inequality $f_\beta(\rho',H')\ge F_\beta'$ applied to the final state $\rho'$ and the final Hamiltonian $H'$.

In particular, if the operation performed on the thermal equilibrium state $\rho_\beta$ is unital, it induces a positive entropy change $\Delta S_\mathrm{unital}(\rho_\beta)\ge0$~\cite{HolevoBook2019,Sagawa2022}, as noted in Sec.~\ref{subsection:NumericalExample1}, and the generalized second law of thermodynamics~(\ref{eq:Generalized2ndLaw}) is reduced to
\begin{equation}
\Delta E_\mathrm{unital}(\rho_\beta)
\ge
\Delta F_\beta.
\label{eq:Unital2ndLaw}
\end{equation}
This reproduces a version of the standard second law of thermodynamics without feedback control.
Note that the bound~(\ref{eq:Unital2ndLaw}) can also be derived from fluctuation relations for unitary operations~\cite{Tasaki2000,Talkner2007,Campisi2011} and for unital operations~\cite{MorikuniTasaki2011,Rastegin2013,Goold2021}.
In the following, we will call the bound~(\ref{eq:Unital2ndLaw}) free-energy bound.
If one performs a nonunital operation, endowed with a feedback structure, on the other hand, it can decrease the entropy, $\Delta S(\rho_\beta)<0$, and the free-energy bound~(\ref{eq:Unital2ndLaw}) can be broken~\cite{MorikuniTasaki2011}.
In other words, feedback control is necessary to break the free-energy bound~(\ref{eq:Unital2ndLaw}).

\subsection{Tightness of the Ergotropy Bound}
\label{subsection:TightnessErgotropyBound}
The energy gain $\Delta E_\mathrm{unital}(\rho_\beta)$ by unital quantum operations from the initial thermal equilibrium state $\rho_\beta$ at a positive inverse temperature $\beta\,(>0)$ is bounded from below by the free-energy bound~(\ref{eq:Unital2ndLaw}).
On the other hand, the same quantity $\Delta E_\mathrm{unital}(\rho_\beta)$ is also bounded from below by the ergotropy bound~(\ref{eq:UnitalErgotropyBound}). 
One can show that the latter is always tighter than the former.

Notice first that the thermal equilibrium state $\rho_\beta$ is a passive state with respect to the initial Hamiltonian $H$, and hence $\mathcal{E}_{\rho_\beta}^-=0$.
Then, the ergotropy bound of~(\ref{eq:UnitalErgotropyBound}) on $\Delta E_\mathrm{unital}(\rho_\beta)$ is reduced to
\begin{equation}
\Delta E_\mathrm{unital}(\rho_\beta)
\ge
\Delta\bm{\varepsilon}^\uparrow\cdot\bm{r}^\downarrow_\beta,
\label{eq:GibbsErgotropyBound}
\end{equation}
where $\bm{r}_\beta^\downarrow$ is the $d$-dimensional vector consisting of the eigenvalues of the density operator $\rho_\beta$ of the initial thermal equilibrium state, arranged in the decreasing order.
This lower bound $\Delta\bm{\varepsilon}^\uparrow\cdot\bm{r}^\downarrow_\beta$ is further bounded from below by $\Delta F_\beta$.
In fact, recalling again $\mathcal{E}_{\rho_\beta}^-=0$ and letting $U^-$ be the optimal unitary achieving $\Delta E_{U^-}(\rho_\beta)=\min_U\Delta E_U(\rho_\beta)$, the ergotropy bound in~(\ref{eq:GeneralizedErgotropyBound}) on the energy gain $\Delta E_U(\rho_\beta)$ by unitary operations yields
\begin{align}
\Delta\bm{\varepsilon}^\uparrow\cdot\bm{r}^\downarrow_\beta
={}&
\min_U\Delta E_{U}(\rho_\beta)
\nonumber\\
={}&
\Delta E_{U^-}(\rho_\beta)
-\frac{1}{\beta}\Delta S_{U^-}(\rho_\beta)
\nonumber\\
\ge{}&
\Delta F_\beta,
\label{eq:FreeEnergyBoundProof}
\end{align}
where the second equality is due to the invariance of the von Neumann entropy under a unitary operation, $\Delta S_U(\rho)=S\bm{(}\Phi_U(\rho)\bm{)}-S(\rho)=0$, and the last inequality is due to the generalized second law of thermodynamics~(\ref{eq:Generalized2ndLaw}), which is based on the inequality~(\ref{eq:FreeEnergiesGap}), namely, in the present case, $f_\beta\bm{(}\Phi_{U^-}(\rho_\beta),H'\bm{)}\ge F_\beta'$.
The equality hence holds if and only if $\Phi_{U^-}(\rho_\beta)=\rme^{-\beta H'}/Z'_\beta$.
We have thus proven that 
\begin{equation}
\Delta E_\mathrm{unital}(\rho_\beta)
\ge
\Delta\bm{\varepsilon}^\uparrow\cdot\bm{r}^\downarrow_\beta
\ge\Delta F_\beta,
\end{equation}
that is, the ergotropy bound of~(\ref{eq:UnitalErgotropyBound}) for the initial thermal equilibrium state $\rho_\beta$ is tighter than the free-energy bound~(\ref{eq:Unital2ndLaw}).
The two bounds coincide if and only if the initial Gibbs state $\rho_\beta$ can be transformed into the Gibbs state $\rme^{-\beta H'}/Z'_\beta$ of the final Hamiltonian $H'$ at the same inverse temperature $\beta$ by the unitary operation $\Phi_{U^-}$ saturating the ergotropy bound for the initial Gibbs state.

Similar inequalities hold for initial Gibbs states $\rho_{-\beta}=\rme^{\beta H}/Z_{-\beta}$ at negative inverse temperatures $-\beta\,(<0)$,
\begin{equation}
\Delta E_\mathrm{unital}(\rho_{-\beta})
\le
\Delta\bm{\varepsilon}^\uparrow\cdot\bm{r}^\uparrow_{-\beta}
\le
\Delta F_{-\beta}
.
\label{eq:NegativeFreeEnergyBound}
\end{equation}
This can be proven by using, instead of the inequality~(\ref{eq:FreeEnergiesGap}),
\begin{equation}
f_{-\beta}(\rho,H)\le F_{-\beta},
\label{eq:FreeEnergiesGapNegative}
\end{equation}
valid for any density operator $\rho$, any Hamiltonian $H$, and any negative inverse temperature $-\beta\,(<0)$.
The flip of the inequality is simply due to the negative sign of the inverse temperature $-\beta\,(<0)$.
See Appendix~\ref{appendix:NegativeFreeEnergiesGap} for its proof.

\subsection{Numerical Examples}
\label{subsection:NumericalExample2}
Let us look at an example. 
We again consider a three-level system, and compare the energy gains $\Delta E(\rho_\beta)$ from the thermal equilibrium states $\rho_\beta$ by unital operations $\Phi_\mathrm{unital}$ and by nonunital operations $\Phi_\mathrm{fb}$. 
Note that in this subsection we consider noncyclic Hamiltonian drives $H\to H'$; otherwise, the {free-energy bound}~(\ref{eq:Unital2ndLaw}) given by the gain in the free energy $\Delta F_\beta$ identically vanishes independent of $\beta$.

\begin{figure}[t]
\begin{tabular}{l@{\ }l}
\footnotesize(a)
&
\footnotesize(b)
\\
\includegraphics[height=0.316\linewidth]{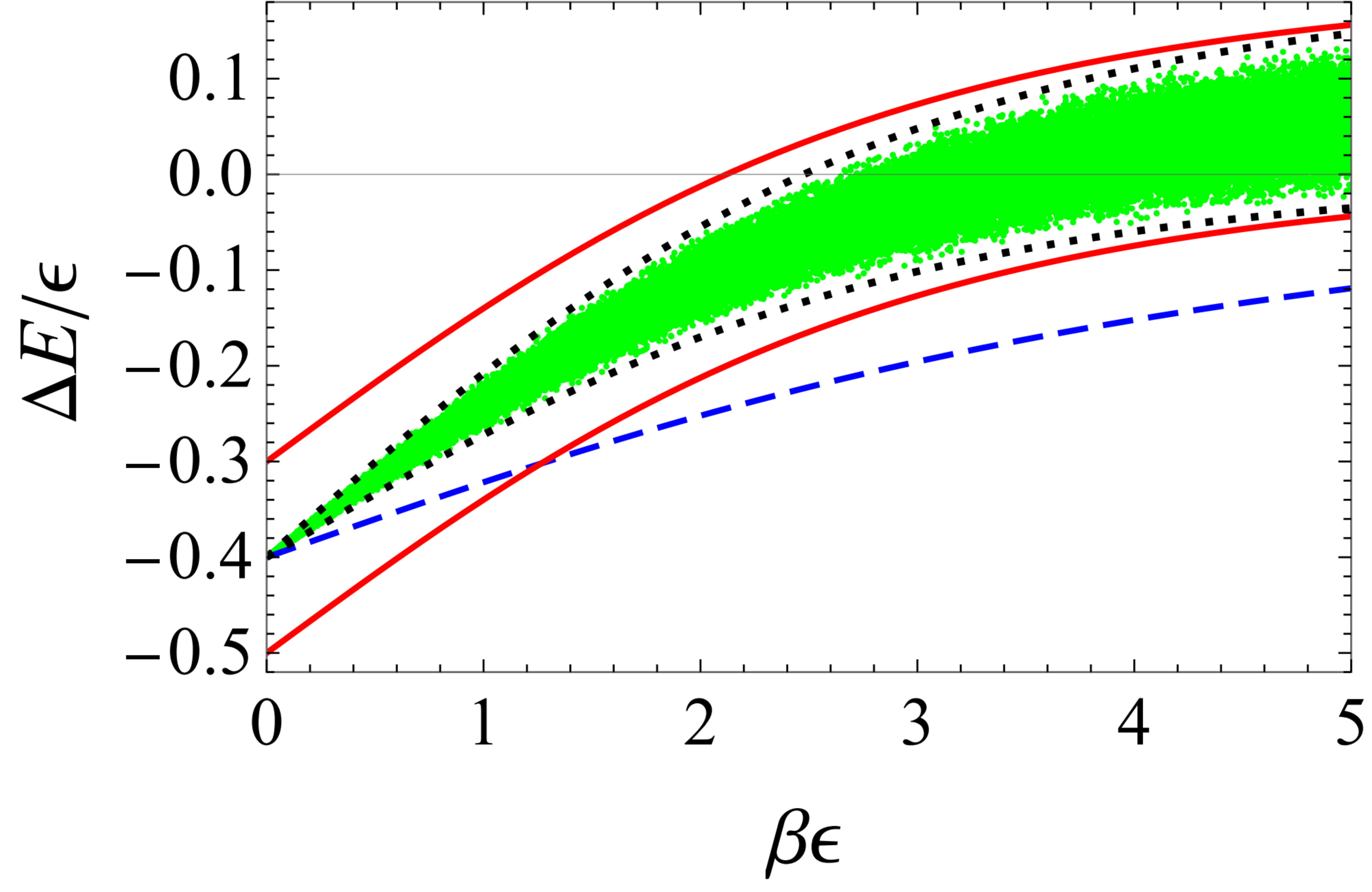}
&
\includegraphics[height=0.316\linewidth]{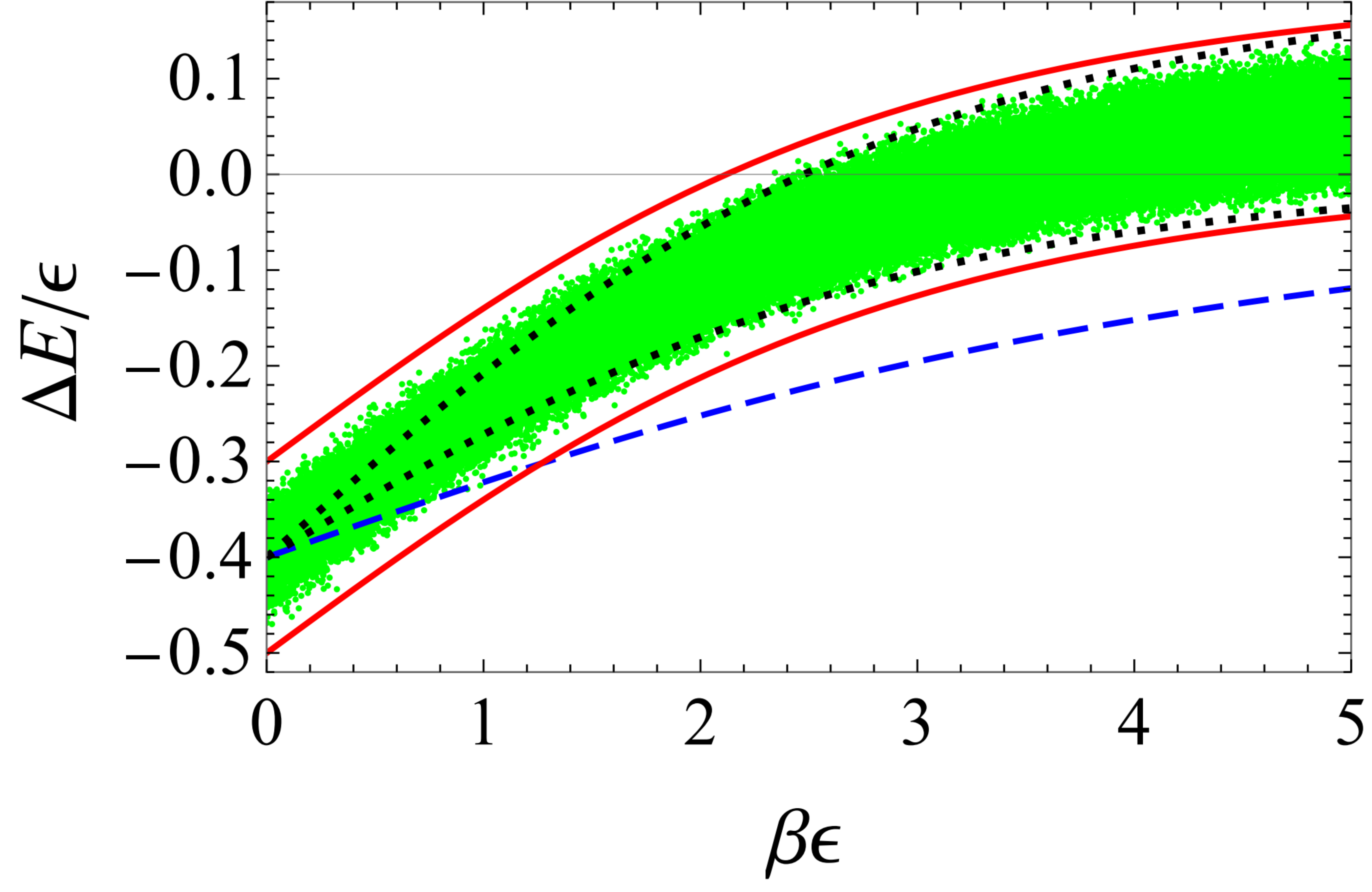}
\end{tabular}
\caption{The energy gains $\Delta E(\rho_\beta)$ from the thermal equilibrium states $\rho_\beta$ of a three-level system by randomly sampled unital operations $\Phi_\mathrm{unital}$ [green dots in (a)] and by randomly sampled nonunital operations $\Phi_\mathrm{fb}$ [green dots in (b)], where the Hamiltonian is driven from the initial one $H=\diag(0,0.5\epsilon,\epsilon)$ to the final one $H'=\diag(0,0.1\epsilon,0.2\epsilon)$.
The unital $\Phi_\mathrm{unital}$ and nonunital $\Phi_\mathrm{fb}$ operations are sampled in the same ways as those in Sec.~\ref{subsection:NumericalExample1}\@.
The dotted lines are the ergotropy and charging bounds~(\ref{eq:UnitalErgotropyBound}) for unital operations $\Phi_\mathrm{unital}$, while the solid red lines are the bounds~(\ref{eq:NonunitalBound}) for general nonunital operations $\Phi_\mathrm{fb}$. 
The dashed blue line indicates the free-energy bound~(\ref{eq:Unital2ndLaw}).}
\label{fig:InitialGibbs1}
\end{figure}

See Fig.~\ref{fig:InitialGibbs1}, where the energy gains $\Delta E(\rho_\beta)$ by randomly sampled unital operations $\Phi_\mathrm{unital}$ and by randomly sampled general nonunital operations $\Phi_\mathrm{fb}$ are shown, together with the ergotropy and charging bounds~(\ref{eq:UnitalErgotropyBound}) for unital operations $\Phi_\mathrm{unital}$ (dotted lines), the bounds~(\ref{eq:NonunitalBound}) for general nonunital operations $\Phi_\mathrm{fb}$ (solid lines), and the free-energy bound~(\ref{eq:Unital2ndLaw}) (dashed line).
Here, the initial and final Hamiltonians are chosen to be $H=\diag(0,0.5\epsilon,\epsilon)$ and $H'=\diag(0,0.1\epsilon,0.2\epsilon)$, respectively.
As shown in the previous subsection, the ergotropy lower bound in~(\ref{eq:UnitalErgotropyBound}) (dotted line) is always tighter than the free-energy bound in~(\ref{eq:Unital2ndLaw}) (dashed line).
One can extract more energy beyond the free-energy bound in~(\ref{eq:Unital2ndLaw}) by some nonunital operations $\Phi_\mathrm{fb}$.
See the data points at low inverse temperatures $\beta$ in Fig.~\ref{fig:InitialGibbs1}(b).
Notice, however, that the free-energy bound in~(\ref{eq:Unital2ndLaw}) (dashed line) can become looser than the lower bound in~(\ref{eq:NonunitalBound}) for general nonunital operations $\Phi_\mathrm{fb}$ (solid line).
In Fig.~\ref{fig:InitialGibbs1}(b), the free-energy bound (dashed line) crosses the lower bound for general nonunital operations (solid line) and becomes looser.
At higher inverse temperatures $\beta$, the free-energy bound in~(\ref{eq:Unital2ndLaw}) (dashed line) cannot be broken even by feedback controls $\Phi_\mathrm{fb}$.

\begin{figure}[t]
\begin{tabular}{l@{\ }l}
\footnotesize(a)
&
\footnotesize(b)
\\
\includegraphics[height=0.312\linewidth]{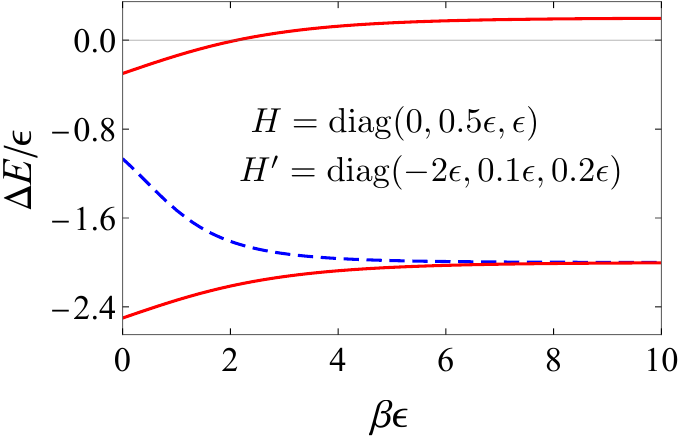}
&
\includegraphics[height=0.312\linewidth]{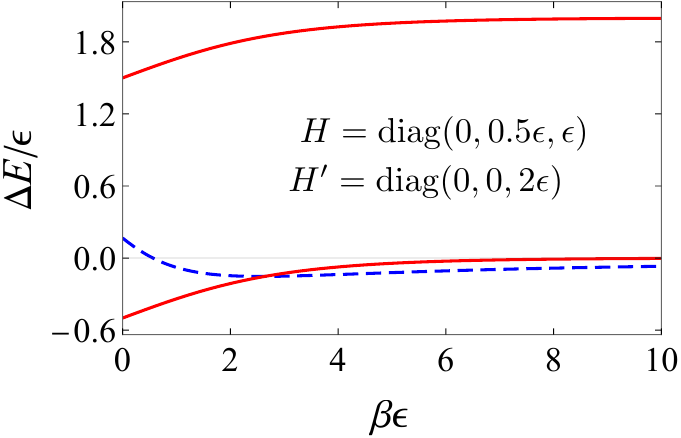}
\\
\footnotesize(c)
&
\footnotesize(d)
\\
\includegraphics[height=0.312\linewidth]{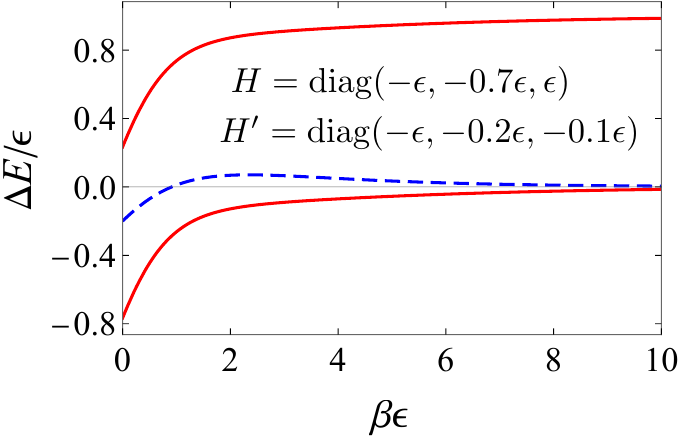}
&
\includegraphics[height=0.312\linewidth]{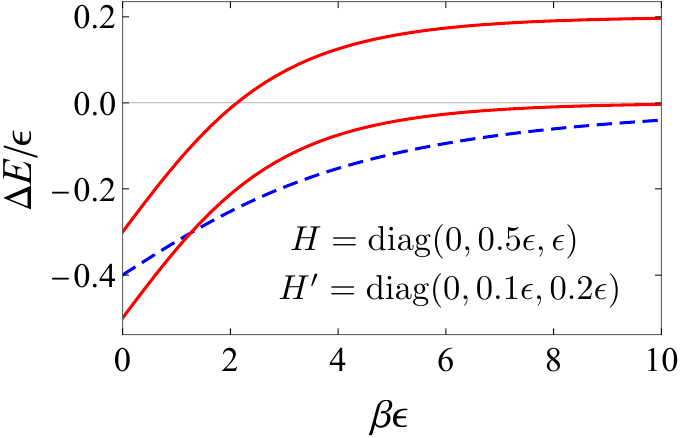}
\\
&
\footnotesize(e)
\\
&
\multicolumn{1}{r}{\includegraphics[height=0.312\linewidth]{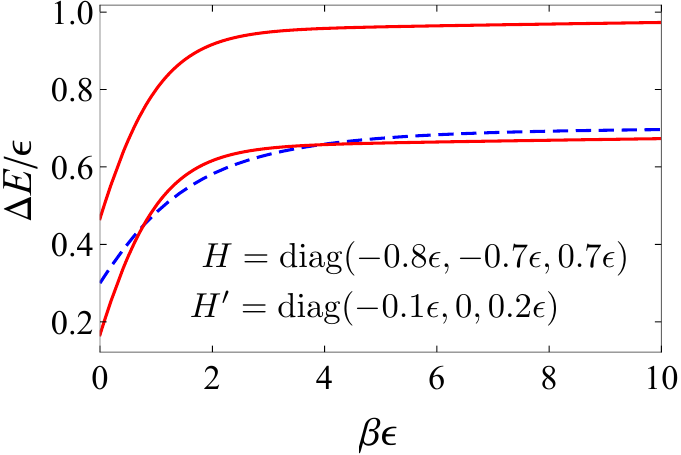}}
\end{tabular}
\caption{Typical patterns of the behaviors of the bounds~(\ref{eq:NonunitalBound}) for general nonunital operations $\Phi_\mathrm{fb}$ (solid red lines) and of the free-energy bound~(\ref{eq:Unital2ndLaw}) (dashed blue line), found for three-level systems.
The bounds~(\ref{eq:NonunitalBound}) for general nonunital operations $\Phi_\mathrm{fb}$ (solid red lines) are monotonically increasing functions of $\beta$, while the free-energy bound~(\ref{eq:Unital2ndLaw}) (dashed blue line) decreases (a)--(b) or increases (c)--(e) in the small $\beta$ region.
The lower bounds (solid red line and dashed blue line) both approach $\varepsilon_1'^\uparrow-\varepsilon_1^\uparrow$ in the limit $\beta\to\infty$. 
They do not cross in (a) and (c), they cross once in (b) and (d), and they cross twice in (e).
}
\label{fig:CrossingPatterns}
\end{figure}

This depends on the choices of the initial and final Hamiltonians $H$ and $H'$.
By randomly sampling the initial and final Hamiltonians $H$ and $H'$ of three-level systems, we typically found the four patterns shown in Figs.~\ref{fig:CrossingPatterns}(a)--(d) for the behaviors of the bounds~(\ref{eq:NonunitalBound}) for general nonunital operations $\Phi_\mathrm{fb}$ (solid lines) and of the free-energy bound~(\ref{eq:Unital2ndLaw}) (dashed line).
For a general $d$-dimensional system, the lower and upper bounds~(\ref{eq:NonunitalBound}) for general nonunital operations $\Phi_\mathrm{fb}$ (solid lines) are monotonically increasing functions of $\beta$ for initial thermal equilibrium states $\rho_\beta$, since the initial energy $E(\rho_\beta)=\Tr(H\rho_\beta)$ is a monotonically decreasing function of $\beta$, approaching the lowest energy $\varepsilon_1^\uparrow$ of the initial Hamiltonian $H$ in the limit $\beta\to\infty$.
On the other hand, the free-energy bound $\Delta F_\beta$ in~(\ref{eq:Unital2ndLaw}) (dashed line) approaches the asymptotic value $\varepsilon_1'^\uparrow-\varepsilon_1^\uparrow$ of the lower bound of~(\ref{eq:NonunitalBound}) for general nonunital operations $\Phi_\mathrm{fb}$ (solid line) in the limit $\beta\to\infty$ monotonically or with a local minimum/maximum, since its derivative $\partial\Delta F_\beta/\partial\beta=\beta^{-2}[S(\rho_\beta')-S(\rho_\beta)]$, where $\rho_\beta'=\rme^{-\beta H'}/Z_\beta'$, can turn its sign if the monotonically decreasing entropies $S(\rho_\beta')$ and $S(\rho_\beta)$ cross.
If the lower bound of~(\ref{eq:NonunitalBound}) for general nonunital operations $\Phi_\mathrm{fb}$ (solid line) crosses the free-energy bound of~(\ref{eq:Unital2ndLaw}) (dashed line), the free-energy bound becomes so loose that any general quantum operations with feedback controls cannot break it at some inverse temperatures $\beta$.
We also found a pattern shown in Fig.~\ref{fig:CrossingPatterns}(e), where the lower bound of~(\ref{eq:NonunitalBound}) for general nonunital operations $\Phi_\mathrm{fb}$ (solid line) crosses the free-energy bound of~(\ref{eq:Unital2ndLaw}) (dashed line) twice.
We do not fully understand the general structure for general $d$-dimensional systems or the physical reason why such crossing occurs, but there certainly exist cases where the free-energy bound~(\ref{eq:Unital2ndLaw}) cannot be broken by any general quantum operations with feedback control.

\begin{figure}[t]
\begin{tabular}{l@{\ }l}
\footnotesize(a)
&
\footnotesize(b)
\\
\includegraphics[height=0.318\linewidth]{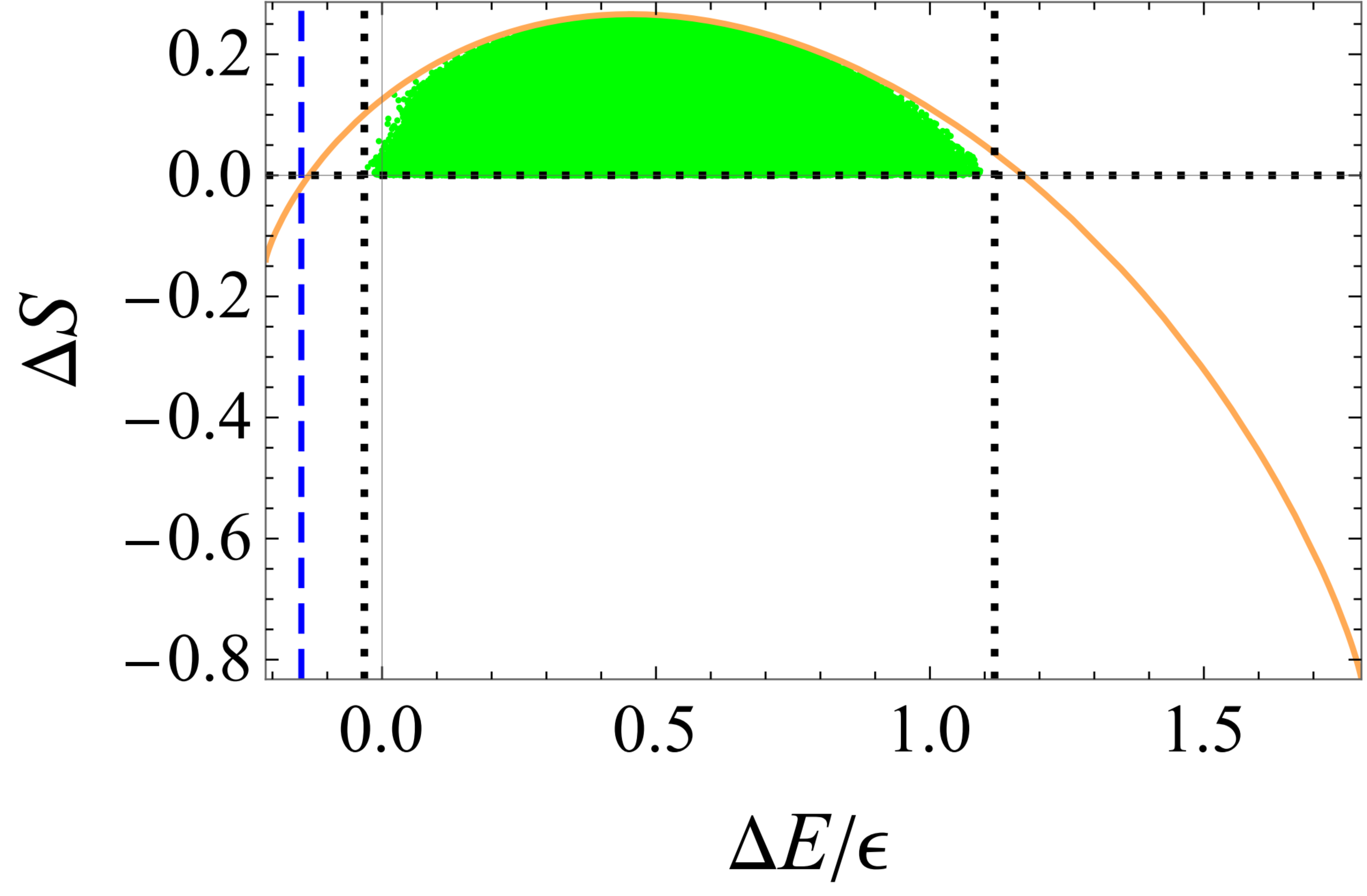}
&
\includegraphics[height=0.318\linewidth]{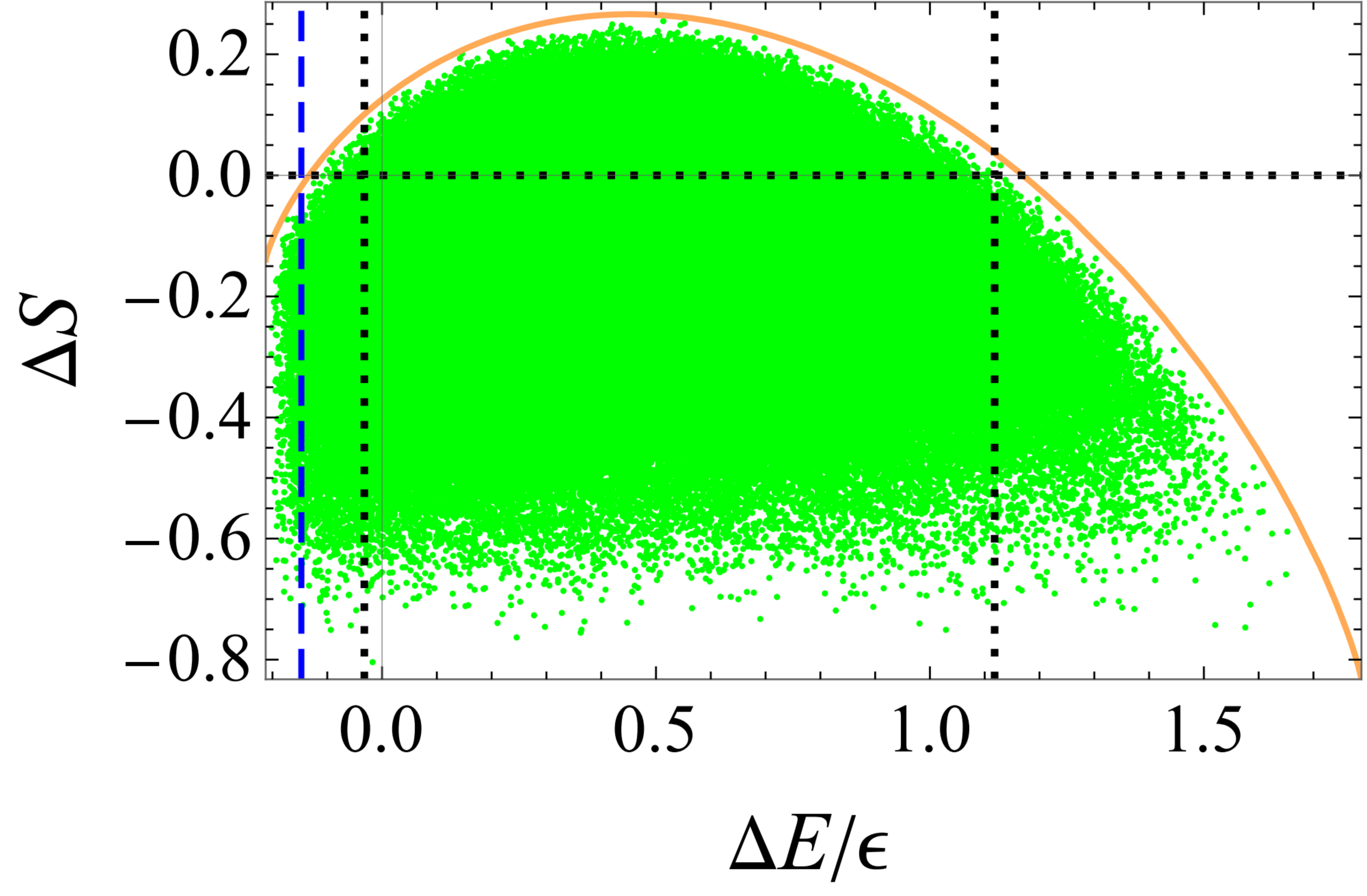}
\end{tabular}
\caption{The energy and entropy gains, $\Delta E(\rho_\beta)$ and $\Delta S(\rho_\beta)$, from the thermal equilibrium state $\rho_\beta$ of a three-level system at the inverse temperature $\beta=2\epsilon^{-1}$ by randomly sampled unital operations $\Phi_\mathrm{unital}$ [green dots in (a)] and by randomly sampled nonunital operations $\Phi_\mathrm{fb}$ [green dots in (b)], where the Hamiltonian is driven from the initial one $H=\diag(0,0.5\epsilon,\epsilon)$ to the final one $H'=\diag(0,0,2\epsilon)$.
The vertical dotted lines are the ergotropy and charging bounds~(\ref{eq:UnitalErgotropyBound}) for unital operations $\Phi_\mathrm{unital}$, while the vertical dashed blue line indicates the free-energy bound~(\ref{eq:Unital2ndLaw}). The solid orange concave curve shows the upper bound on the allowed entropy gain $\Delta S(\rho_\beta)$. The horizontal dotted line indicates $\Delta S(\rho_\beta)=0$.}
\label{fig:InitialGibbs2}
\end{figure}

Finally, as mentioned below~(\ref{eq:Unital2ndLaw}), it is necessary to decrease the entropy, $\Delta S(\rho_\beta)<0$, to break the free-energy bound~(\ref{eq:Unital2ndLaw}).
To break the ergotropy bound~(\ref{eq:UnitalErgotropyBound}), which is tighter than the free-energy bound~(\ref{eq:Unital2ndLaw}), it is not necessarily the case.
See Fig.~\ref{fig:InitialGibbs2}.
Here, the initial and final Hamiltonians are chosen to be $H=\diag(0,0.5\epsilon,\epsilon)$ and $H'=\diag(0,0,2\epsilon)$, respectively, and the inverse temperature is set at $\beta=2\epsilon$.
We see in Fig.~\ref{fig:InitialGibbs2}(b) that there exist nonunital operations $\Phi_\mathrm{fb}$ that increase the entropy, $\Delta S(\rho_\beta)>0$, and at the same time break the ergotropy bound~(\ref{eq:UnitalErgotropyBound}) (see the data points above the horizontal dotted line indicating the equientropic line and on the left of the vertical dotted line indicating the ergotropy lower bound)~\cite{footnote:TwoLevelSystem}.
In order to break the free-energy bound~(\ref{eq:Unital2ndLaw}), on the other hand, it is necessary to decrease the entropy, $\Delta S(\rho_\beta)<0$ (see the data points on the left of the vertical dashed line indicating the free-energy bound).

\section{conclusions}
\label{section:Conclusion}
We have studied the energy extraction and charging by general quantum operations, in particular focusing on the roles of quantum measurement and quantum feedback control.
We have shown that the energy gain by a unital quantum operation is bounded by the ergotropy/charging bound for unitary operations.
This implies that, in order to extract/charge energy beyond the ergotropy/charging bound for unitary operations, feedback control is necessary.
We have also shown that the ergotropy/charging bound proven here for unital operations is tighter than the standard second law of thermodynamics.

In Ref.~\cite{Solfanelli2019}, the maximum extractable energy by projective measurement is studied, and it is shown to be strictly smaller than the ergotropy.
It appears that, if one performs quantum measurement but does not apply any feedback control, the ergotropy/charging bound~(\ref{eq:UnitalErgotropyBound}) for unital quantum operations is not saturable.
It would be intriguing to explore the tight bound on the energy gain by general quantum measurement, to acquire better understanding on the possible roles and the limitations of quantum measurement in quantum thermodynamics.

\acknowledgments
KY acknowledges supports by the Top Global University Project from the Ministry of Education, Culture, Sports, Science and Technology (MEXT), Japan, and by the Grants-in-Aid for Scientific Research (C) (No.~18K03470) and for Fostering Joint International Research (B) (No.~18KK0073) both from the Japan Society for the Promotion of Science (JSPS).

\appendix
\section{Proof of the Representation~(\ref{eq:UnitalIsNoFeedback}) of Unital Map}
\label{appendix:UnitalIsNoFeedback}
We here provide the proof of the decomposition of unital map $\Phi_\mathrm{unital}$ in~(\ref{eq:UnitalIsNoFeedback}).
Let $\rho=\sum_{i=1}^dr_i|r_i\rangle\langle r_i|$ be a density operator in the diagonalized form, and $\Phi_\mathrm{unital}(\rho)=\sum_{i=1}^dr_i'|r_i'\rangle\langle r_i'|$ be the eigenvalue decomposition of the output state of a unital map $\Phi_\mathrm{unital}(\rho)$. 
It is known that the output state $\Phi_\mathrm{unital}(\rho)$ of a unital map is majorized by the input state $\rho$, i.e., $\rho\succ\Phi_\mathrm{unital}(\rho)$, or equivalently $\bm{r}\succ\bm{r}'$, which means $\sum_{i=1}^kr_i^\downarrow\ge\sum_{i=1}^kr_i^{{\prime}{\downarrow}}$, for all $k=1,\ldots,d$ (Theorem~6.1 of Ref.~\cite{Sagawa2022}). 
Then, the Schur-Horn theorem ensures that there exists a Hermitian matrix $X$ whose eigenvalues and diagonal elements are given by the elements of $\bm{r}$ and $\bm{r}'$, respectively (Theorem~4.3.48 of Ref.~\cite{HornJohnson2012} and Theorem~9.B.2 of Ref.~\cite{Marshall2011}). 
Since $X$ is diagonalizable with a unitary matrix $V$ as $X=V\diag(r_1,\ldots,r_d)V^\dag$, the diagonal elements $X_{ii}=r_i'$ satisfy
\begin{equation}
r_i'
=
\sum_{j=1}^d|V_{ij}|^2 r_j.
\end{equation}
Now, let us take a complete set of orthonormal basis vectors $\{\ket{\psi_i}\}$ fulfilling $V_{ij}=\langle\psi_i|r_j\rangle$, and construct orthogonal projections $P_i=|\psi_i\rangle\langle \psi_i|$.
In addition, let $U$ be the unitary connecting the basis vectors as $U|\psi_i\rangle=|r_i'\rangle$.
Then, these elements yield
\begin{align}
\sum_{i=1}^d
UP_i\rho P_iU^\dag
={}&
\sum_{i=1}^d\sum_{j=1}^d
U|\psi_i\rangle
\langle\psi_i|r_j\rangle r_j\langle r_j|\psi_i\rangle
\langle\psi_i|U^\dag
\nonumber
\displaybreak[0]
\\
={}&
\sum_{i=1}^d\sum_{j=1}^d
|V_{ij}|^2 r_j|r_i'\rangle\langle r_i'|
\nonumber
\displaybreak[0]
\\
={}&
\sum_{i=1}^dr_i'|r_i'\rangle\langle r_i'|
\nonumber
\displaybreak[0]
\\
={}&
\Phi_\mathrm{unital}(\rho),
\vphantom{\sum_{i=1}^d}
\end{align}
which proves the representation~(\ref{eq:UnitalIsNoFeedback}).
Note that the projectors $\{P_i\}$ and the unitary $U$ depend on the input state $\rho$.

\section{Proof of the Generalized Ergotropy Bound~(\ref{eq:GeneralizedErgotropyBound})}
\label{appendix:GeneralizedErgotropyBound}
In this appendix, we prove the generalized ergotropy bound~(\ref{eq:GeneralizedErgotropyBound}) for unitary operations $\Phi_U$.
It suffices to prove
\begin{equation} 
\bm{\varepsilon}^{\prime\uparrow}
\cdot
\bm{r}^\downarrow
\le
\Tr(H' U\rho U^\dag)
\le
\bm{\varepsilon}^{\prime\uparrow}
\cdot
\bm{r}^\uparrow.
\label{eq:EquivalentErgotropyBound}
\end{equation}
We prove it in basically the same way as the one used for the proof of passivity on $d$-dimensional quantum systems~\cite{Lenard1978,ThirringBook2002,Koukoulekidis2021}.
See also Theorem~4.3.53 of Ref.~\cite{HornJohnson2012}.
We first observe that
\begin{align}
\Tr(H' U\rho U^\dag)
={}&
\sum_{i=1}^d\sum_{j=1}^d
\varepsilon_i'|\langle\varepsilon_i'|U|r_j\rangle|^2
r_j
\nonumber\\
={}&
\bm{\varepsilon}'\cdot B\bm{r},
\end{align}
where $|\varepsilon_i'\rangle$ and $|r_j\rangle$ are the eigenstates belonging to the eigenvalues $\varepsilon_i'$ and $r_j$ of $H'$ and $\rho$, respectively, and $B_{ij}=|\langle\varepsilon_i'|U|r_j\rangle|^2$ is a doubly stochastic (or bistochastic) matrix, satisfying $B_{ij}\ge0$ and $\sum_{j=1}^dB_{ij}=\sum_{j=1}^dB_{ji}=1$ for all $i=1,\ldots,d$. 
The Birkhoff-von Neumann theorem (Theorem~8.7.2 of Ref.~\cite{HornJohnson2012} and Theorem~2.A.2 of Ref.~\cite{Marshall2011}) ensures that there exists a probability distribution $\{p_k\}_{k=1,\ldots,d!}$ such that
\begin{equation}
B=\sum_{k=1}^{d!}p_k\Pi_k,
\label{eq:BirkhoffvonNeumann}
\end{equation}
with $\{\Pi_k\}_{k=1,\ldots,d!}$ the set of permutation matrices permuting the $d$ elements of $d$-dimensional vectors.
In addition, we recall the rearrangement inequality (Proposition~6.A.3 of Ref.~\cite{Marshall2011} and Theorem~10.4 of Ref.~\cite{Zhang2011})
\begin{equation}
\bm{x}^\uparrow\cdot\bm{y}^\downarrow
\le
\bm{x}\cdot\Pi_k\bm{y}
\le
\bm{x}^\uparrow\cdot\bm{y}^\uparrow,
\label{eq:PermInnerProd}
\end{equation}
for any pair of $d$-dimensional real vectors $\bm{x}$ and $\bm{y}$, and for all permutations $k=1,\ldots,d!$.
Then, using its upper bound, we get
\begin{align}
\Tr(H'U\rho U^\dag)
={}&
\sum_{k=1}^{d!}p_k\bm{\varepsilon}'\cdot\Pi_k\bm{r}
\nonumber
\displaybreak[0]
\\
\le{}&
\sum_{k=1}^{d!}p_k\bm{\varepsilon}^{\prime\uparrow}\cdot\bm{r}^\uparrow
\nonumber
\\
={}&
\bm{\varepsilon}^{\prime\uparrow}\cdot\bm{r}^\uparrow,
\end{align}
which is the upper bound of~(\ref{eq:EquivalentErgotropyBound}).
The lower bound of~(\ref{eq:EquivalentErgotropyBound}) is proven by using the lower bound of~(\ref{eq:PermInnerProd}).

Both upper and lower bounds of~(\ref{eq:EquivalentErgotropyBound}) are reached by some unitaries $U^+$ and $U^-$, respectively.
Indeed, $U^{-(+)}=\sum_{i=1}^d|\varepsilon_i^{\prime\uparrow}\rangle\langle r_i^{\downarrow(\uparrow)}|$, which transforms the initial state $\rho$ into the passive~\cite{Pusz1978,Lenard1978} (maximally active~\cite{Binder2015NJP}) state $\rho_{P(A)}=\sum_{i=1}^dr_i^{\downarrow(\uparrow)}|\varepsilon_i^{\prime\uparrow}\rangle\langle\varepsilon_i^{\prime\uparrow}|$ with respect to the final Hamiltonian $H'$, does the job.

Note that the bound~(\ref{eq:EquivalentErgotropyBound}) holds for arbitrary unitary $U$ and Hamiltonian $H'$.
By using~(\ref{eq:EquivalentErgotropyBound}) for $U=\openone$ and $H'=H$, the positivities $\mathcal{E}_\rho^-\ge0$ and $\mathcal{E}_\rho^+\ge0$ of (\ref{eq:ErgotropyMinus}) and (\ref{eq:ErgotropyPlus}) are proven.

\section{Proof of the Inequality~(\ref{eq:FreeEnergiesGapNegative}) for the Nonequilibrium Free Energy at a Negative Temperature}
\label{appendix:NegativeFreeEnergiesGap}
The inequality~(\ref{eq:FreeEnergiesGap}) at a positive inverse temperature $\beta>0$ is a fundamental relation between the equilibrium and nonequilibrium free energies, $F_\beta$ and $f_\beta(\rho,H)$~\cite{Landi2021,Esposito2011,Sagawa2022}. 
We here prove the similar inequality~(\ref{eq:FreeEnergiesGapNegative}) at a negative inverse temperature $-\beta<0$. 
As in the case of a positive inverse temperature $\beta>0$, it is due to the positivity of the quantum relative entropy $D(\rho\|\sigma)=\Tr[\rho(\log\rho-\log\sigma)]\ge0$, whose equality holds if and only if $\rho=\sigma$~\cite{Nielsen2010,Hayashi2015,Sagawa2022}. 
For the Gibbs state $\rho_{-\beta}=\rme^{\beta H}/Z_{-\beta}$ at a negative inverse temperature $-\beta<0$, we have
\begin{align}
0
\ge{}&
{-\frac{1}{\beta}}D(\rho\|\rho_{-\beta})
\nonumber
\\
={}&
{-\frac{1}{\beta}}\Tr(\rho\log\rho)+\frac{1}{\beta}\Tr(\rho\log\rho_{-\beta})
\nonumber\\
={}&
\frac{1}{\beta}S(\rho)+\Tr(H\rho)-\frac{1}{\beta}\log Z_{-\beta}
\nonumber\\
={}&
f_{-\beta}(\rho,H)-F_{-\beta},
\vphantom{\frac{1}{\beta}}
\end{align}
which proves~(\ref{eq:FreeEnergiesGapNegative}).
The equality $f_{-\beta}(\rho,H)=F_{-\beta}$ holds if and only if $\rho=\rho_{-\beta}$.


\begin{thebibliography}{10}
\bibitem{footnote:UnitalFeedback}
Note that the projections $\{P_i\}$ and the unitary $U$ depend also on the input state $\rho$.
\bibitem{footnote:UnitalSpaces}
The region is not filled by the dots. As already noted, the ergotropy and charging bounds in~(\ref{eq:UnitalErgotropyBound}) can be reached by some unitary operations $\Phi_U$, and hence, the points along the equientropic line between the ergotropy and charging bounds in Fig.~\ref{fig:ESdiagram1}(a) are reachable. The tip of the concave curve in Fig.~\ref{fig:ESdiagram1}(a) corresponds to the maximally mixed state $\rho_0=\openone/3$ with $\beta=0$, and it is also reachable from any state by the completely depolarizing operation, which is unital. On the other hand, there are spaces that are not reachable by unital operations $\Phi_\mathrm{unital}$, i.e., the regions around the points where the concave curve intersects with the ergotropy and charging lines.
\bibitem{footnote:WholeRegion}
Indeed, there exists a CPTP map that brings a system from any initial state to the Gibbs state $\rho_\beta$ with an arbitrary $\beta\in\mathbb{R}$. Such a CPTP map is not unital and is endowed with a feedback structure. We then perform a unitary operation to bring the system from the Gibbs state $\rho_\beta$ to the target state along the equientropic line. Any target state lies within the ergotropy or charging range from a Gibbs state with a positive or negative $\beta$.
\bibitem{footnote:TwoLevelSystem}
This happens only for systems with more than two levels. For two-level systems, it is necessary to decrease the entropy to break the ergotropy bound~(\ref{eq:UnitalErgotropyBound}). See Fig.~\ref{fig:ESdiagram4}(b).
\end{thebibliography}

\end{document}